\documentclass[acmsmall]{acmart}

\acmJournal{PACMPL}
\acmVolume{1}
\acmNumber{CONF} 
\acmArticle{1}
\acmYear{2018}
\acmMonth{1}
\acmDOI{} 
\startPage{1}

\setcopyright{none}

\usepackage{booktabs}   
\usepackage{subcaption} 

\usepackage{kotex}
\usepackage{fancyvrb}
\usepackage{nameref}
\usepackage{hyperref}
\usepackage{xspace}
\usepackage{cleveref}
  \Crefformat{chapter}{\S#2#1#3}
  \Crefformat{section}{\S#2#1#3}
  \Crefformat{figure}{Fig. #2#1#3}
\usepackage{iris}
\usepackage{wrapfig}
\usepackage{lipsum}
\usepackage{fancyvrb}
\usepackage{balance}
\usepackage{mathpartir}
\usepackage{stmaryrd}
\usepackage{cprotect}
\usepackage{pbox}
\usepackage[normalem]{ulem}
\usepackage{dashbox}
\usepackage{soul}
\usepackage{xcolor}
\usepackage{coqdoc}
\usepackage{multirow}
\usepackage{framed}
\usepackage{wasysym}
\usepackage{tabu}

\definecolor{varpurple}{rgb}{0.4,0,0.4}
\definecolor{constrmaroon}{rgb}{0.6,0,0}
\definecolor{defgreen}{rgb}{0,0.4,0}
\definecolor{indblue}{rgb}{0,0,0.8}
\definecolor{kwred}{rgb}{0.8,0.1,0.1}
\definecolor{yjpurple}{rgb}{0.25,0,0.25}
\definecolor{yjgreen}{rgb}{0.25,0.5,0.25}

\def\coqdocvarcolor{yjpurple}
\def\coqdockwcolor{yjgreen}
\def\coqdoccstcolor{defgreen}
\def\coqdocindcolor{indblue}
\def\coqdocconstrcolor{constrmaroon}
\def\coqdocmodcolor{defgreen}
\def\coqdocaxcolor{varpurple}
\def\coqdoctaccolor{black}

\def\coqdockw#1{{\color{\coqdockwcolor}{{\texttt{#1}}}}}
\def\coqdocvar#1{{\color{\coqdocvarcolor}{\hspace{1pt}\texttt{#1}\hspace{1pt}}}}
\def\coqdoccst#1{{\color{\coqdoccstcolor}{\textrm{#1}}}}
\def\coqdocind#1{{\color{\coqdocindcolor}{\textsf{#1}}}}
\def\coqdocconstr#1{{\color{\coqdocconstrcolor}{\textsf{#1}}}}
\def\coqdocmod#1{{{\color{\coqdocmodcolor}{\textsc{\textsf{#1}}}}}}
\def\coqdocax#1{{{\color{\coqdocaxcolor}{\textsl{\textrm{#1}}}}}}

\begin{document}

\sethlcolor{lightgray}

\title[Abstraction Logic]{Abstraction Logic: The Marriage of Contextual Refinement and Separation Logic}         


\author{Youngju Song} \affiliation{\institution{Seoul National University} \country{Korea}} \email{youngju.song@sf.snu.ac.kr}
\author{Minki Cho}    \affiliation{\institution{Seoul National University} \country{Korea}} \email{minki.cho@sf.snu.ac.kr}
\author{Dongjae Lee}  \affiliation{\institution{Seoul National University} \country{Korea}} \email{dongjae.lee@sf.snu.ac.kr}
\author{Chung-Kil Hur}\affiliation{\institution{Seoul National University} \country{Korea}} \email{gil.hur@sf.snu.ac.kr}

\newcommand{\ie}{\textit{i.e.,} }
\newcommand{\cf}{\textit{cf.} }
\newcommand{\eg}{\textit{e.g.,} }
\newcommand{\etal}{\textit{et al.}}

\newcommand{\gil}[1]{}
\newcommand{\yj}[1]{}
\newcommand{\minki}[1]{}
\newcommand{\ldj}[1]{}
\newcommand{\mytitle}[1]{}
\newcommand{\todo}[1]{}


\newcommand{\code}[1]{\texttt{#1}\xspace}
\newcommand{\valr}{r}
\newcommand{\beh}[1]{\text{Beh}({#1})}
\newcommand{\crems}{AL\xspace}
\newcommand{\CREMS}{abstraction logic\xspace}
\newcommand{\ems}{\textrm{EMS}\xspace}
\newcommand{\imp}{IMP\xspace}
\newcommand{\spc}{SPC\xspace}
\newcommand{\gen}{\textrm{PAbs}\xspace}
\newcommand{\CR}{CR}
\newcommand{\checker}{SpecChecker\xspace}
\newcommand{\barr}{\noindent\makebox[\linewidth]{\rule{\textwidth}{0.4pt}}}
\newcommand{\exitval}{0}
\newcommand{\asm}[1]{\textcolor{red}{#1}}
\newcommand{\grt}[1]{\textcolor{blue}{#1}}
\newcommand{\link}{\circ}
\newcommand{\iris}{Iris\xspace}
\newcommand{\vst}{VST\xspace}
\newcommand{\lectx}{\le_{\text{ctx}}}
\newcommand{\boilerplate}[1]{\textcolor{white!60!black}{#1}}
\newcommand{\mires}{\sigma}

\newcommand{\hide}[1]{}
\newcommand{\unhide}[1]{#1}

\newcommand{\myparagraph}[1]{\paragraph{\hspace*{-3.8mm}\bfseries{#1}}}
\definecolor{grey}{rgb}{0.5,0.5,0.5}
\newcommand{\caselabel}[1]{\begin{small}{\color{grey}{(\textsc{#1})}}\end{small}}
\newcommand{\mycomment}[1]{\begin{small}{\vspace{1.5mm}\color{grey}{(*****\;#1\;*****)}}\end{small}}

\newcommand{\twoA}{{2\textrm{A}}}
\newcommand{\twoB}{{2\textrm{B}}}

\newenvironment{stackAux}[2]{%
  \setlength{\arraycolsep}{0pt}
  \begin{array}[#1]{#2}}{
  \end{array}}
\newenvironment{stackCC}{
  \begin{stackAux}{c}{c}}{\end{stackAux}}
\newenvironment{stackCL}{
  \begin{stackAux}{c}{l}}{\end{stackAux}}
\newenvironment{stackTL}{
  \begin{stackAux}{t}{l}}{\end{stackAux}}
\newenvironment{stackTR}{
  \begin{stackAux}{t}{r}}{\end{stackAux}}
\newenvironment{stackBC}{
  \begin{stackAux}{b}{c}}{\end{stackAux}}
\newenvironment{stackBL}{
  \begin{stackAux}{b}{l}}{\end{stackAux}}

\NewDocumentCommand \ahoare {m m m O{}}{
	\curlybracket{#1}& \spac #2 \spac &\curlybracket{#3}_{#4}%
}

\newcommand{\impl}{{\textsl{impl}}}
\newcommand{\spec}{{\textsl{spec}}}
\newcommand{\inimpl}[1]{\ensuremath{#1_\impl}}
\newcommand{\inspec}[1]{\ensuremath{#1_\spec}}

\newcommand{\cfbox}[2]{%
    \colorlet{currentcolor}{.}%
    {\color{#1}%
    \fbox{\color{currentcolor}#2}}%
}
\newcommand{\cdashbox}[2]{%
    \colorlet{currentcolor}{.}%
    {\color{#1}%
    \dashbox{\color{currentcolor}#2}}%
}
\newcommand{\defeq}{\ensuremath{\stackrel{\text{def}}{=}}}
\newcommand{\defcoind}{\ensuremath{\stackrel{\text{coind}}{=}}}

\newcommand{\echoimpl}{\text{$I_\text{\code{Echo}}$}\xspace}
\newcommand{\echospec}{\text{$S_\text{\code{Echo}}$}\xspace}
\newcommand{\echoabs}{\text{$A_\text{\code{Echo}}$}\xspace}
\newcommand{\echogen}{\text{$G_\text{\code{Echo}}$}\xspace}
\newcommand{\stackimpl}{\text{$I_\text{\code{Stack}}$}\xspace}
\newcommand{\stackspec}{\text{$S_\text{\code{Stack}}$}\xspace}
\newcommand{\stackabs}{\text{$A_\text{\code{Stack}}$}\xspace}
\newcommand{\stackgen}{\text{$G_\text{\code{Stack}}$}\xspace}
\newcommand{\memimpl}{\text{$I_\text{\code{Mem}}$}\xspace}
\newcommand{\memspec}{\text{$S_\text{\code{Mem}}$}\xspace}
\newcommand{\memabs}{\text{$A_\text{\code{Mem}}$}\xspace}
\newcommand{\memtop}{\text{$\overline{A}_\text{\code{Mem}}$}\xspace}
\newcommand{\memgen}{\text{$G_\text{\code{Mem}}$}\xspace}
\newcommand{\bwimpl}{\text{$I_\text{\code{Bw}}$}\xspace}
\newcommand{\bwspec}{\text{$S_\text{\code{Bw}}$}\xspace}
\newcommand{\bwgen}{\text{$G_\text{\code{Bw}}$}\xspace}

\newcommand{\kw}[1]{{\code{\textbf{#1}}}}
\newcommand{\kwdef}{\kw{def}}
\newcommand{\kwspec}{\kw{spec}}
\newcommand{\kwfriend}{\kw{friend}}
\newcommand{\kwcontext}{\kw{context}}
\newcommand{\kwvar}{\kw{var}}
\newcommand{\kwif}{\kw{if}}
\newcommand{\kwthen}{\kw{then}}
\newcommand{\kwelse}{\kw{else}}
\newcommand{\kwmatch}{\kw{match}}
\newcommand{\kwwith}{\kw{with}}
\newcommand{\kwend}{\kw{end}}
\newcommand{\kwassume}{{\color{red}\kw{assume}}}
\newcommand{\kwguarantee}{{\color{blue}\kw{guarantee}}}
\newcommand{\kwgrnt}{{\color{blue}\kw{grnt}}}
\newcommand{\ASSUME}{\textcolor{red}{\code{ASSUME}}}
\newcommand{\GUARANTEE}{\textcolor{blue}{\code{GUARANTEE}}}
\definecolor{darkblue}{rgb}{0.0, 0.0, 0.55}
\definecolor{darkred}{rgb}{0.55, 0.0, 0.0}
\newcommand{\kwupdate}{\ensuremath{\rightsquigarrow=}}
\newcommand{\kwaddp}{\kw{\hspace{0.3mm}+=}}
\newcommand{\kwsubp}{\kw{\hspace{0.3mm}-=}}
\newcommand{\kwmodule}{\kw{Module}}
\newcommand{\kwdefault}{\kw{default}}
\newcommand{\kwskip}{\kw{skip}}
\newcommand{\kwfpu}{\kw{fpu}}
\newcommand{\kwminus}{\kw{minus}}

\newcommand{\kwchoose}{\kw{choose}}
\newcommand{\kwtake}{\kw{take}}
\newcommand{\kwsyscall}{\kw{syscall}}
\newcommand{\kwloop}{\kw{loop}}
\newcommand{\kwbreak}{\kw{break}}
\newcommand{\kwcall}{\kw{call}}
\newcommand{\kwlocal}{\kw{local}}
\newcommand{\kwub}{\kw{UB}}
\newcommand{\kwnb}{\kw{NB}}
\newcommand{\kwfun}{\kw{fun}}
\newcommand{\kwwhile}{\kw{while}}

\newcommand{\res}{\code{res}}
\newcommand{\resframe}{\code{res\_frame}}
\newcommand{\kwfp}{\code{res\_f}}
\newcommand{\kwmp}{\code{res\_m}}
\newcommand{\kwcanonpcm}{\code{$\textrm{PCM}_\Cannon$}}

\newcommand{\kwmalloc}{\kw{malloc}}
\newcommand{\kwfree}{\kw{free}}
\newcommand{\kwload}{\kw{load}}
\newcommand{\kwstore}{\kw{store}}
\newcommand{\kwrepeat}{\kw{repeat}}

\newcommand{\kwpure}{\textrm{TRIVIAL}}

\newcommand{\Snone}{S_{*}}
\newcommand{\snone}{s_{*}}

\newcommand{\optionm}{\ensuremath{\textdom{Option}}}

\newcommand{\fsem}{f_\textrm{sem}}
\newcommand{\Sf}{S_\code{f}}
\newcommand{\RP}{\code{RP}}
\newcommand{\SC}{\code{SC}}
\newcommand{\AD}{\code{AD}}
\newcommand{\IO}{\code{IO}}
\newcommand{\Echo}{\code{Echo}}
\newcommand{\Mem}{\code{Mem}}
\newcommand{\Stack}{\code{Stack}}
\newcommand{\Main}{\code{Main}}
\newcommand{\main}{\code{main}}
\newcommand{\MF}{\code{F}}
\newcommand{\mf}{\code{f}}
\newcommand{\Cannon}{\code{Cannon}}
\newcommand{\fire}{\code{fire}}
\newcommand{\kwundef}{\code{Undef}}
\newcommand{\kwemp}{\code{Unit}}
\newcommand{\ball}{\code{Ball}}
\newcommand{\slball}{\ownGhost{}{\code{Ball}}}
\newcommand{\ready}{\code{Ready}}
\newcommand{\fired}{\code{Fired}} 
\newcommand{\numfire}{\code{NUM\_FIRE}}
\newcommand{\powder}{\code{powder}}

\newcommand{\madd}{+}
\newcommand{\nullptr}{\code{NULL}}

\newcommand{\usable}[1]{\ownGhost{}{\textrm{Expect(#1)}}}
\newcommand{\usablex}[1]{\textrm{Expect(#1)}}
\newcommand{\usablexnoarg}[1]{\textrm{Expect}}
\newcommand{\being}[1]{\ownGhost{}{\textrm{Indeed(#1)}}}
\newcommand{\beingx}[1]{\textrm{Indeed(#1)}}
\newcommand{\beingxnoarg}{\textrm{Indeed}}
\newcommand{\full}[1]{\ownGhost{}{\textrm{Ok(#1)}}}
\newcommand{\fullx}[1]{\textrm{Ok(#1)}}
\newcommand{\fullxnoarg}{\textrm{Ok}}
\newcommand{\unitt}{\textrm{Emp}}
\newcommand{\wrong}{\textrm{Wrong}}
\newcommand{\true}{\textrm{T}}
\newcommand{\false}{\textrm{F}}
\newcommand{\colorcode}{\textrm{hex}}
\newcommand{\bb}{\texttt{b}}
\newcommand{\wdef}{\mval}

\newcommand{\prefire}{\code{pre\_fire}}
\newcommand{\postfire}{\code{post\_fire}}
\newcommand{\premain}{\code{pre\_main}}
\newcommand{\postmain}{\code{post\_main}}

\newcommand{\name}{\code{name}}

\newcommand{\downcast}[1]{{\ensuremath{{\downarrow}{#1}}}}
\newcommand{\upcast}[1]{{\ensuremath{{\uparrow}{#1}}}}
\newcommand{\unwrapU}{{\color{red}?}}
\newcommand{\unwrapN}{{\color{blue}!}}
\newcommand{\empval}{\code{ev}}

\newcommand{\sljudge}[3]{{#1 \model #2 : #3}}
\newcommand{\spcmod}[3]{{[#1 \rtimes #2 : #3]}}
\newcommand{\safe}[1]{{#1 \text{ safe}}}
\newcommand{\Safe}{\text{Safe}\,}
\newcommand{\spcabs}[1]{{#1^\text{A}}}
\newcommand{\divcode}{\,\code{/}\,}
\newcommand{\modcode}{\,\code{\%}\,}
\newcommand{\prt}{$print\hspace*{-.5pt}$}
\newcommand{\cmt}[1]{\textcolor{white!60!black}{//#1}}

\newcommand{\mydashbox}[3]{\hspace*{1pt}\textcolor{#1}{\dashbox{\parbox{#2}{#3}}}}
\setlength{\dashdash}{1pt}
\setlength{\dashlength}{2pt}

\newcommand{\option}{\code{option}}
\newcommand{\vunit}{\code{()}}
\newcommand{\vlist}{\code{list}}
\newcommand{\lsthead}{\textrm{head}}
\newcommand{\lsttail}{\textrm{tail}}
\newcommand{\lstnz}{\textrm{nonzero}}
\newcommand{\lstlength}{\textrm{length}}
\newcommand{\some}{\code{Some}}
\newcommand{\none}{\code{None}}
\newcommand{\vval}{\code{val}}
\newcommand{\vptr}{\code{ptr}}
\newcommand{\vint}{\code{int64}}
\newcommand{\vbool}{\code{bool}}
\newcommand{\vundef}{\code{vundef}}
\newcommand{\rplift}[1]{\lc #1 \rc}
\newcommand{\lc}{\ulcorner}
\newcommand{\rc}{\urcorner}
\newcommand{\own}[1]{\ownGhost{}{#1}}
\newcommand{\opure}{\some}
\newcommand{\otop}{\none}
\newcommand{\lrepeat}{\textrm{repeat}}
\newcommand{\blk}{\textrm{b}}
\newcommand{\ofs}{\textrm{ofs}}
\newcommand{\xp}{\textrm{p}}
\newcommand{\xv}{\textrm{v}}
\newcommand{\xx}{\textrm{x}}
\newcommand{\xr}{\textrm{r}}
\newcommand{\handle}{h}
\newcommand{\stack}{\textrm{stk}}
\newcommand{\stackz}{\textrm{estk}}

\newcommand{\argp}{x}
\newcommand{\argv}{x_\textrm{a}}
\newcommand{\argo}{d}
\newcommand{\retp}{r}
\newcommand{\retv}{r_\textrm{a}}
\newcommand{\lst}{\ell}

\newcommand{\tyany}{\anyty}
\newcommand{\tyt}{\textrm{T}}
\newcommand{\mypoint}[1]{\textbf{\emph{(#1)}}}
\newcommand{\isstack}[2]{{\textrm{is\_stk}\;#1\;#2}}
\newcommand{\isbag}[2]{{\textrm{is\_bag}\;#1\;#2}}
\newcommand{\isestack}[2]{{\textrm{is\_estk}\;#1\;#2}}
\newcommand{\rom}[1]{\uppercase\expandafter{\romannumeral #1\relax}}
\newcommand{\myset}[2]{(#2 \Rightarrow #1)}

\newcommand{\llink}{\link}
\newcommand{\llinkall}{\ensuremath{\fullmoon}}
\newcommand{\plink}{\ensuremath{\bullet}}
\newcommand{\trm}[1]{\textrm{#1}}

\newcommand{\lland}{\,\land\,}
\newcommand{\emsmod}{\trm{Mod}}
\newcommand{\emsmods}{\trm{Mods}}
\newcommand{\LD}{\trm{LD}}
\newcommand{\mlift}{\trm{lift}}

\newcommand{\anyty}{\code{Any}}

\newcommand{\toabs}[2]{[#1]_{#2}}
\newcommand{\setofz}[1]{\{#1\}}
\newcommand{\metaspec}[1]{\overline{#1}}

\newcommand{\kwgetcaller}{\kw{get\_caller}}

\newcommand{\world}{\ensuremath{\mathcal{W}}}
\newcommand{\simrel}{I}
\newcommand{\transA}{toAbs}
\newcommand{\transAS}{toAbspec}
\newcommand{\AS}{abspec}


\definecolor{cverbbg}{gray}{0.93}

\newenvironment{lcverbatim}
 {\SaveVerbatim{cverb}}
 {\endSaveVerbatim
  \flushleft\fboxrule=0pt\fboxsep=.5em
  \colorbox{cverbbg}{%
    \makebox[\dimexpr\linewidth-2\fboxsep][l]{\BUseVerbatim{cverb}}%
  }
  \endflushleft
}

\definecolor{shadecolor}{gray}{0.93}
\newcommand{\mbind}{\ensuremath{>\!\!>\!=}}
\newcommand{\mcat}{\ensuremath{>\!\!>\!\!>}}

\newcommand{\rProp}{\mathbf{rProp}}

\newcommand{\ordinal}{\code{ord}}

\newcommand{\Meta}{\code{A}}
\newcommand{\meta}{a}
\newcommand{\cnd}{\code{c}}

\DeclareRobustCommand\model{\mathrel{|}\joinrel\mkern-.5mu\mathrel{\textrm{--}}}


\newcommand{\ord}{ord}
\newcommand{\anyv}{Any}
\newcommand{\propb}{\mathbb{P}}
\newcommand{\typeb}{Type}

\newcommand{\effectarrow}{\hookrightarrow}

\newcommand{\itree}[2]{itree #1 #2}

\newcommand{\fname}{string}
\newcommand{\mname}{string} 

\newcommand{\chooseE}{\code{Choose}}
\newcommand{\takeE}{\code{Take}}
\newcommand{\syscallE}{\code{Obs}}

\newcommand{\getNameE}{\code{GetCaller}}
\newcommand{\putE}{\code{Put}}
\newcommand{\getE}{\code{Get}}
\newcommand{\callE}{\code{Call}}
\newcommand{\ApcE}{\code{IPC}}

\newcommand{\EventE}{\ensuremath{\trm{E}_{\text{prim}}}}
\newcommand{\ModE}{\ensuremath{\trm{E}_{\text{mod}}}}
\newcommand{\EmsE}{\ensuremath{\trm{E}_{\text{\ems}}}}
\newcommand{\SpcE}{\ensuremath{\trm{E}_{\text{\gen}}}}

\newcommand{\unitset}{()}

\newcommand{\SLtriple}{SL\_triple}
\newcommand{\SLspec}{SL\_spec}
\newcommand{\SLspecs}{SL\_spectable}

\newcommand{\mfun}[1]{function#1}

\newcommand{\spcmodforget}[2]{|#1|$_{\text{#2}}$}

\newcommand{\spcfun}[4]{{\{#1 \rtimes #2 : #3 / #4\}}}
\newcommand{\spcfunforget}[3]{|#1 / #2|\texorpdfstring{\textsubscript{#3}}}

\newcommand{\SLfun}{SL\_decorate}

\newcommand{\SLpre}{SL\_arg}
\newcommand{\SLpost}{SL\_ret}
\newcommand{\SLcall}{SL\_call}
\newcommand{\SLbody}{SL\_body}
\newcommand{\SLIPC}{SL\_IPC}
\newcommand{\SLsafe}{SL\_safe}

\newcommand{\rprop}{rProp}

\newcommand{\all}{g}
\newcommand{\names}{ns}

\newcommand{\simle}{\ensuremath{\sqsubseteq_{_{\world}}}}
\newcommand{\ievent}{effect}

\newcommand{\kwrp}{\code{res\_r}}
\newcommand{\kwap}{\code{res\_x}}

\newcommand{\iside}{\code{i}}
\newcommand{\aside}{\code{a}}

\newcommand{\myhrule}{\\[2mm]\hrule}
\newcommand{\mynewpage}{\newpage\noindent}

\newcommand{\fn}{f\!n}
\newcommand{\frm}{\trm{frm}}

\newcommand{\xorv}{xr}



\begin{abstract}
  Contextual refinement and separation logics are successful verification techniques that are very different in nature.
First, the former guarantees behavioral refinement between a concrete program and an abstract program
while the latter guarantees safety of a concrete program under certain conditions (expressed in terms of pre and post conditions).
Second, the former does not allow any assumption about the context when locally reasoning about a module while the latter allows rich assumptions.

In this paper, we present a new verification technique, called abstraction logic (AL), that inherently combines contextual refinement and separation logics such as Iris and VST, thereby taking the advantages of both.
Specifically, AL allows us to locally verify a concrete module against an abstract module under separation-logic-style pre and post conditions about external modules.
AL are fully formalized in Coq and provides a proof mode that supports a combination of simulation-style reasoning using our own tactics and SL-style reasoning
using IPM (Iris Proof Mode).
Using the proof mode, we verified various examples to demonstrate reasoning about ownership (based on partial commutative monoids) and purity ($i.e.$, termination with no system call), cyclic and higher-order reasoning about mutual recursion and function pointers, and reusable and gradual verification via intermediate abstractions.
Also, the verification results are combined with CompCert, so that we formally establish behavioral refinement from top-level abstract programs, all the way down to their assembly code.



\end{abstract}

\ccsdesc[500]{Software and its engineering~General programming languages}
\ccsdesc[300]{Social and professional topics~History of programming languages}


\maketitle

%

\section{Introduction}
\label{sec:introduction}

Contextual refinement~\cite{gu:dscal} and program logics~\cite{hoare1969logic} (most notably, separation logics~\cite{reynolds2002separation, ohearn2007csl}) are two successful verification techniques.
The former is typically used for compiler verification, where both implementation and specification are given as executable programs and its verification establishes that all possible observable behaviors of the implementation program under an arbitrary context are included in those of the specification program under the same context.
The latter are mainly used for verification of a program against its logical specification, where the specification is given as a pair of pre and post conditions and its verification typically establishes that if the program starts with a state satisfying the pre condition, then it executes safely, and if it terminates, the final state satisfies the post condition.

These two techniques are very different in nature and have their own advantages.
First, specifications in contextual refinement describe intended dynamic behaviors including interactions with its context/environment and side effects such as fatal errors and non-termination, while those in program logics typically describe sufficient conditions for \emph{safe} executions (\ie those without fatal errors in case of partial correctness, and in addition without non-termination in case of total correctness).
Second, contextual refinement allows to verify one aspect at a time via multiple intermediate (specification) programs since they are transitively composable (\eg a compiler translation consists of a sequence of optimizations, each of which is separately verified using a possibly different simulation relation), while program logics require to find and verify safety conditions for the whole program in one step (although the conditions may be refined in further steps).
Third, when locally reasoning about a module, in contextual refinement one cannot make any assumptions about external modules because they are completely arbitrary,
while in program logics one can make specific assumptions (\ie safety conditions) about each external module since the verified module is only composed with other \emph{verified} modules instead of \emph{arbitrary} modules.

In this paper, we present a new verification framework, called \CREMS (\crems), that \emph{inherently} combines the two techniques of contextual refinement~(CR) and separation logic~(SL), thereby taking the advantages of both.

\paragraph{\textbf{High-level overview of \crems}}
We first highlight how \crems is different from CR and the standard version of SL using an abstract example (See \Cref{sec:related} for comparison with other variations of SL).
For this, consider two modules, named $\code{M}$ with a function \code{f} and named $\code{N}$ with a function \code{g},
and suppose we have their implementations $I_1$, $I_2$ and SL-style specifications $S_1$, $S_2$.

In SL, one can \emph{locally} verify each implementation $I_i$ for $i\in \{1,2\}$ against its specification $S_i$ assuming both specifications $S_1,S_2$,
which we denote by $\sljudge{S_1,S_2}{I_i}{S_i}$. Then SL combines the two verification results as follows:
\[
  \inferrule{\sljudge{S_1,S_2}{I_1}{S_1} \\\\ \sljudge{S_1,S_2}{I_2}{S_2}}{\safe{I_1 \link I_2}}
\]
It guarantees that the linked program $I_1 \link I_2$ only produces safe behaviors (\ie no fatal errors).
For example, if $\code{M}.\code{f}() \equiv (1+1)\divcode\code{N}.\code{g}()$ in $I_1$, then by assuming $\code{N}.\code{g}()$ is safe and returns $1$, which is specified in $S_2$, we can prove that $\code{M}.\code{f}()$ is safe and returns $2$, which is specified in $S_1$.
It is important to note that SL in general cannot guarantee anything when verified modules are composed with unverified ones because those unverified may trigger a fatal error.
For example, if $\code{M}.\code{f}() \equiv (\code{L}.\code{h}(); (1+1)\divcode\code{N}.\code{g}())$ in $I_1$ and the implementation $I_3$ for the module $\code{L}$ is unverified, SL cannot guarantee safety of $I_1 \link I_2 \link I_3$ since $I_3$ might not be safe.

On the other hand, CR provides verification results that are valid under unverified contexts.
Specifically, in CR, one can \emph{locally}\footnote{
  In theory, it might be possible to \emph{globally} verify $I_1 \link I_2$, where you can make assumptions about $I_1$ and $I_2$.
  However, it would sacrifice the power of local reasoning and compositionality, so we aim high to support fully compositional verification.
}
verify $I_1$ against another more abstract implementation $A_1$, simply called \emph{abstraction}, for $\code{M}$, which we denote by $I_1 \lectx A_1$.
Then by the definition of CR, we have the following result for an arbitrary (even unsafe or abstract) implementation $C$ for $\code{N}$.
\[
\inferrule{I_1 \lectx A_1}{\beh{I_1\link C} \subseteq \beh{A_1 \link C}}
\]
For example, when $\code{M}.\code{f}() \equiv (1+1)\divcode\code{N}.\code{g}()$ in $I_1$,
we can verify it against $A_1$ with $\code{M}.\code{f}() \equiv 2\divcode\code{N}.\code{g}()$,
which is valid even when \code{N} is an unverified arbitrary module
because the behavioral refinement preserves even crash or non-termination behaviors between the implementation and abstraction (\eg if the implementation crashes, so does the abstraction).
However, its limitation is that we cannot verify $I_1$ against more useful abstractions such as
$\code{M}.\code{f}() \equiv 2$ because $\code{N}.\code{g}()$ in $C$ is arbitrary and thus may not return $1$.

The key innovation in \crems is that we overcome the limitation of CR by internalizing, inside a module, SL-style specifications about other modules.
To see this, revisit the above problematic example where $\code{M}.\code{f}() \equiv (1+1)\divcode\code{N}.\code{g}()$ in $I_1$ and $\code{M}.\code{f}() \equiv 2$ in $A_1$ and thus $I_1 \lectx A_1$ does not hold.
To solve this problem, from $A_1$ together with the above specifications $S_1$, $S_2$ (\ie saying that $\code{M}.\code{f}()$ returns $2$ and $\code{N}.\code{g}()$ returns $1$),
\crems derives a special abstraction for $\code{M}$, denoted $\spcmod{S_1,S_2}{A_1}{S_1}$ and called \emph{abspec},
that internalizes $S_1$, $S_2$ inside $A_1$.
Then, one can actually prove $I_1 \lectx \spcmod{S_1,S_2}{A_1}{S_1}$, which essentially amounts to proving that the behaviors of $I_1$ refines those of $A_1$ and satisfies $S_1$ under arbitrary contexts but assuming the specifications $S_1$ and $S_2$.
Note that this verification is \emph{local} to $I_1$ and $A_1$ for $\code{M}$ just relying on the specifications $S_1$ and $S_2$,
so that the refinement can hold under an arbitrary implementation $C$ for $\code{N}$ (as required by the definition of $\lectx$).
Even further, when $\code{M}.\code{f}() \equiv (\code{L}.\code{h}(); (1+1)\divcode\code{N}.\code{g}())$ in $I_1$ and $\code{M}.\code{f}() \equiv (\code{L}.\code{h}(); 2)$ in $A_1$ with the implementation of \code{L} \emph{unknown} (\ie \code{L} may even be unsafe or make arbitrary calls including mutually recursive calls to the modules \code{M} and \code{N}),
we can still prove $I_1 \lectx \spcmod{S_1,S_2}{A_1}{S_1}$ for $S_1, S_2$ the same as above without assuming anything about $\code{L}$.

To combine local verification results, \crems provides the following theorem for any $S_1,S_2,A_1,A_2$.
\[
\spcmod{S_1,S_2}{A_1}{S_1} \link \spcmod{S_1,S_2}{A_2}{S_2} \lectx A_1 \link A_2
\]
Then, by horizontal and vertical compositionality of CR and the definition of CR, we can derive, \eg the following corollary for any (even unsafe or abstract) implementation $C$ for \code{L}:
\[
\inferrule{I_1 \lectx \spcmod{S_1,S_2}{A_1}{S_1} \\\\ I_2 \lectx \spcmod{S_1,S_2}{A_2}{S_2}}
          {\beh{I_1 \link I_2 \link C} \subseteq \beh{A_1 \link A_2 \link C}}
\]

We have two advantages from the fact that the resulting abstraction $A_1 \link A_2 \link C$ is also an executable program.
\begin{enumerate}
\item One can test the abstraction by executing it, so that we can more easily see whether the abstraction works as intended.
\item One can treat the abstraction as an implementation and further verify it using \crems, which allows gradual abstraction from an implementation to the top-level abstraction in multiple steps and can also increase reusability of verification (See \Cref{sec:advanced-gradual} for a concrete example).
An example of such gradual and modular verification is depicted as follows.\\[1mm]
\mbox{}\hfill\includegraphics[scale=.4]{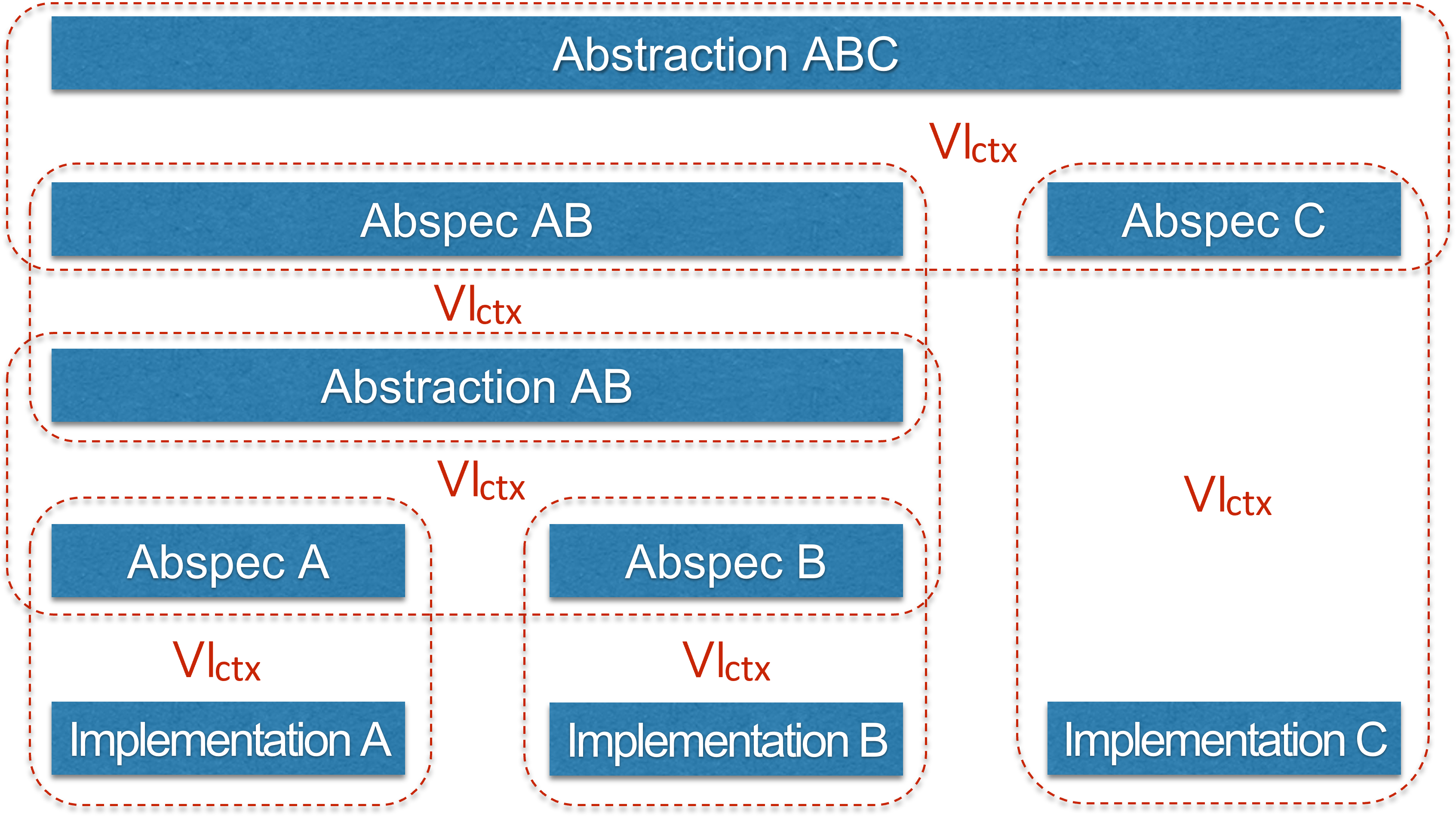}\hfill\mbox{}

\end{enumerate}

Finally, we remark that \crems can be seen as subsuming both CR and SL.
The former is achieved by not making any assumptions,
and the latter because \crems can also be used to prove safety guarantees as follows.
For example, for any $I_1, I_2$ and $S_1,S_2$,
by setting $A_i$ to be a special abstraction, called $\text{Safe}$, we have the following:
\[
\inferrule{I_1 \lectx \spcmod{S_1,S_2}{\text{Safe}}{S_1}  \\\\  I_2 \lectx \spcmod{S_1,S_2}{\text{Safe}}{S_2}}
          {\beh{I_1 \link I_2} \subseteq \beh{\text{Safe} \link \text{Safe}}}
\]
where proving $I_i \lectx \spcmod{S_1,S_2}{\text{Safe}}{S_i}$ essentially amounts to proving $\sljudge{S_1,S_2}{I_i}{S_i}$ in SL.
Then since $\beh{\text{Safe} \link \text{Safe}}$ is a set of safe behaviors,
we can conclude that $I_1 \link I_2$ only produces safe behaviors.

\paragraph{\textbf{Contributions}}
In this paper, we developed the theory of \CREMS (currently in a sequential setting) and tools for it including the \crems proof mode and a verified compiler for \crems down to assembly. All results are fully formalized in Coq~\cite{coq} and summarized as follows.
\begin{enumerate}
\item We developed the first \emph{comprehensive} theory, \crems, that enables establishing contextual refinement
  via powerful local reasoning that allows us to rely on SL-style specifications.
  In particular, \crems allows us to express various ownership via PCMs (Partial Commutative Monoids)~\cite{calcagno2007local} as in the state-of-the-art SLs such as \iris~\cite{iris2015} and \vst~\cite{vst}.
\item We developed \ems~(Executable Module Semantics) by generalizing Interaction Trees~\cite{liyao:itree} to support module systems and two kinds of nondeterminism, called \emph{choose} and \emph{take}.
  In particular, via Coq's extraction mechanism, we can extract an \ems to an executable program in OCaml as done in Interaction Trees.
  Also, we developed two sub-languages \imp and \spc that are embedded into \ems (via deep embedding for \imp and shallow embedding for \spc).
  \begin{itemize}
  \item \imp: a C-like language with integer and (function and memory) pointer values, which is used to write underlying implementations like $I_1$ above that are verified against higher-level abstractions using \crems and also compiled down to assembly via our verified compiler for \imp.
  \item \spc: a specification language, which is used to write an abspec,
    which is a combination of an abstraction and SL-style specifications like $\spcmod{S_1,S_2}{A_1}{S_1}$ above.
  \end{itemize}
\item We developed the \crems proof mode that supports a combination of simulation-style reasoning and SL-style reasoning by allowing smooth switching between the two styles during a single proof, which is essentially necessary because \crems inherently combines the two techniques.
Specifically, for the former we developed our own tactics that allow to set up a simulation relation (possibly with a module-local relational invariant) between the implementation and abspec, and reason about it stepwise;
and for the latter we employed the IPM (\iris Proof Mode) package (\ie by instantiating it with our \crems theory) to streamline the process for reasoning about separating conjunction, magic wand, and PCM resources.

\item Using the \crems proof mode, we verified various examples written in \imp, which demonstrates reasoning about PCM-based ownership, proving and exploiting \emph{purity} (\ie termination with no system call), cyclic and higher-order reasoning about recursion and function pointers, and reusable and gradual verification via intermediate abstractions.
  Note that in spite of supporting cyclic and higher-order reasoning, \crems does not rely on any step-indexing techniques~\cite{ahmed2006step}.

\item Finally, we developed a \emph{verified} compiler for \imp targeting Csharpminor of CompCert~\cite{CompCert}, which is verified in the style of CompCert's verification.
  Therefore, we formally establish behavioral refinement from the top-level abstractions of the above examples, all the way down to their assembly code generated by the \imp compiler and CompCert.
\end{enumerate}


\section{Key ideas}
\label{sec:overview}

We gradually introduce the key ideas behind \crems{} by presenting how to express Hoare logic
specifications (\Cref{sec:overview:hoare}) and separation logic specifications
(\Cref{sec:overview:separation}).


\subsection{Hoare logic specifications in \CREMS{}}
\label{sec:overview:hoare}

To demonstrate how to express Hoare logic~(HL) specifications in \CREMS,
consider the implementations $I_\code{Main}$ and $I_\code{F}$ of the modules \Main{} and \MF{}, shown in \Cref{fig:hoare-logic}.
Here \code{Main.main()} $(i)$ invokes \code{F.f(x)} with $\code{x} = 40$, which computes \code{x*x/4+x+1}, prints it out via the system call \prt{} and returns it;
$(ii)$ if the result is an odd number, prints \code{42}; $(iii)$ otherwise, crashes by executing division by zero.
In HL, the specification $S_\Main$ says that \code{Main.main()} runs safely, \emph{assuming} the specification $S_\MF$,
which says that if the argument of \code{F.f} is a multiple of \code{4}, it runs safely and returns a number $r$ such that $r$ modulus \code{4} is {1}.

In \crems, assuming the specifications $S_\code{Main}$ and $S_\code{F}$ (ignoring the safety), we would like to \emph{locally} verify $I_\code{Main}$ and $I_\code{F}$ against, \eg the abstractions $A_\code{Main}$ and $A_\code{F}$ in \Cref{fig:hoare-logic}, where the abstracted parts are \fbox{boxed} and \code{**} is the exponentiation operator.
First, note that without any assumption neither $I_\code{Main} \lectx A_\code{Main}$ nor $I_\code{F} \lectx A_\code{F}$ holds
because their context implementations are arbitrary.
For example, \code{0} may be given for \code{r} in $\code{Main.main()}$, and \code{1} may be given for \code{x} in $\code{F.f(x)}$, in which cases the implementation and abstraction behave differently.
The question here is how to \emph{internalize} the specifications inside $A_\code{Main}$ and $A_\code{F}$.

\begin{figure}[t]
$I_\code{Main}$ :=
\begin{minipage}[t]{0.4\textwidth}
\begin{Verbatim}[commandchars=\\\{\},codes={\catcode`$=3},fontsize=\small]
[\kwmodule Main]
\kwdef main() $\equiv$
  \kwvar x := 40;
  \kwvar r := F.f(x);
  \kwif (r\%2 == 1) \kwthen \prt(42) \kwelse 1/0
\end{Verbatim}
\end{minipage}%
\hfill%
$I_\code{F}$ :=
\begin{minipage}[t]{0.39\textwidth}
\begin{Verbatim}[commandchars=\\\{\},codes={\catcode`$=3},fontsize=\small]
[\kwmodule F]
\kwdef f(x) $\equiv$
  \kwvar r := x*x/4 + x + 1;
  \prt(r);
  r
\end{Verbatim}
\end{minipage}%
\\[1mm]
\begin{minipage}[t]{0.51\textwidth}
$S_\code{Main}$ := $\setofz{ \hoare{\TRUE} {\code{Main.main}} {\TRUE} }$
\end{minipage}%
\hfill%
\begin{minipage}[t]{0.44\textwidth}
$S_\code{F}$ := $\setofz{ \hoare{\lambda x.\; x \modcode 4 = 0} {\code{F.f}} {\lambda \valr.\; \valr \modcode 4 = 1} }$
\end{minipage}%
\\[1mm]
$A_\Main$ :=
\begin{minipage}[t]{0.4\textwidth}
\begin{Verbatim}[commandchars=\\\{\},codes={\catcode`$=3},fontsize=\small]
[\kwmodule Main]
\kwdef main() $\equiv$
  \kwvar x := 40;
  \kwvar r := F.f(x);
  \fbox{\prt(42)}
\end{Verbatim}
\end{minipage}%
\hfill%
$A_\code{F}$ :=
\begin{minipage}[t]{0.38\textwidth}
\begin{Verbatim}[commandchars=\\\{\},codes={\catcode`$=3},fontsize=\small]
[\kwmodule F]
\kwdef f(x) $\equiv$
  \kwvar r := \fbox{(x/2 + 1)**2};
  \prt(r);
  r
\end{Verbatim}
\end{minipage}%
\\[1mm]
\begin{minipage}[t]{0.5\textwidth}
$\spcmod{S_\code{Main} \cup S_\code{F}}{A_\code{Main}}{S_\code{Main}}$ :=\\
\hspace*{10pt}\begin{minipage}[t]{0.4\textwidth}  
\begin{Verbatim}[commandchars=\\\{\},codes={\catcode`$=3},fontsize=\small]
[\kwmodule Main]
\kwdef main() $\equiv$
  \kwvar x := 40;
 \mydashbox{blue}{10pc}{\kwguarantee(x\%4 == 0);}
  \kwvar r := F.f(x);
 \mydashbox{red}{10pc}{\kwassume(r\%4 == 1);}
  \prt(42)
\end{Verbatim}
\end{minipage}%
\end{minipage}%
\hfill%
\begin{minipage}[t]{0.445\textwidth}
$\spcmod{S_\code{Main} \cup S_\code{F}}{A_\code{F}}{S_\code{F}}$ :=\\
\hspace*{10pt}\begin{minipage}[t]{0.4\textwidth}  
\begin{Verbatim}[commandchars=\\\{\},codes={\catcode`$=3},fontsize=\small]
[\kwmodule F]
\kwdef f(x: int) $\equiv$
 \mydashbox{red}{10pc}{\kwassume(x\%4 == 0);}
  \kwvar r := (x/2 + 1)**2;
  \prt(r);
 \mydashbox{blue}{10pc}{\kwguarantee(r\%4 == 1);}
  r
\end{Verbatim}
\end{minipage}%
\end{minipage}%
\myhrule
\caption{Implementations, HL specifications, Abstractions, and Abspecs for \Main{} and \MF{}}
\label{fig:hoare-logic}
\end{figure}



The abspecs $\spcmod{S_\code{Main} \cup S_\code{F}}{A_\code{Main}}{S_\code{Main}}$ and $\spcmod{S_\code{Main} \cup S_\code{F}}{A_\code{F}}{S_\code{F}}$
internalizing $S_\code{Main},S_\code{F}$ inside $A_\code{Main}$ and $A_\code{F}$ are given in \Cref{fig:hoare-logic},
where we simply insert \kwassume{} and \kwguarantee{} commands according to $S_\code{Main}$ and $S_\code{F}$.
Specifically, \code{F.f(x)} \emph{assumes} its precondition \code{x\%4 == 0} at the beginning and \emph{guarantees} its postcondition \code{r\%4 == 1} at the end.
Conversely, \code{Main.main()} \emph{guarantees} the precondition of \code{F.f(x)} before invoking it and
\emph{assumes} the postcondition of \code{F.f(x)} after the invocation.
Note that we simply omitted \code{\kwassume(true)} and \code{\kwguarantee(true)} corresponding to $\hoare{\TRUE}{}{\TRUE}$ of \code{Main.main()}
because they do nothing as we will see below.

Now the computational interpretation of \kwassume{} and \kwguarantee{} is given as follows.
First, $\kwassume(P)$ for a proposition $P$ does nothing if $P$ holds; otherwise triggers \emph{undefined behavior (UB)}, which is a standard notion (\eg in CompCert) and interpreted as exhibiting all possible behaviors:
\[
\kwassume(P) ~\defeq~ \code{\kwif{} $P$ \kwthen{} \kwskip{} \kwelse{} \kwub{}}
\]
To see the intuition, consider proving $I_\code{F} \lectx \spcmod{S_\code{Main} \cup S_\code{F}}{A_\code{F}}{S_\code{F}}$.
If the argument \code{x} to \code{F.f} does not satisfy \code{x\%4 == 0}, the abspec triggers \kwub{} thereby exhibiting all possible behaviors, which trivially include whatever behavior the implementation may exhibit.
Therefore in the verification we do not need to consider the cases where the assume command fails (\ie we can assume it holds).

Second, $\kwguarantee(P)$ does nothing if $P$ holds; otherwise triggers \emph{no behavior (NB)}, which appeared in~\cite{compcerttso, int2ptr} and is interpreted as exhibiting no behaviors (see \Cref{sec:formal} for the formal definition):
\[
\kwguarantee(P) ~\defeq~ \code{\kwif{} $P$ \kwthen{} \kwskip{} \kwelse{} \kwnb{}}
\]
Again, for $I_\code{F} \lectx \spcmod{S_\code{Main} \cup S_\code{F}}{A_\code{F}}{S_\code{F}}$, suppose the argument \code{x} satisfies \code{x\%4 == 0}
and then both implementation and abspec will compute and print the same value \code{r}.
Now if \code{r} does not satisfy \code{r\%4 == 1}, the abspec triggers \kwnb{} thereby exhibiting no behaviors, which does not include whatever behavior the implementation may exhibit (unless it also triggers \kwnb{}, which is not the case here).
Therefore in the verification we must prove that the guarantee command succeeds (\ie we should guarantee it holds).

Then, we can actually prove $I_\code{Main} \lectx \spcmod{S_\code{Main} \cup S_\code{F}}{A_\code{Main}}{S_\code{Main}}$
and $I_\code{F} \lectx \spcmod{S_\code{Main} \cup S_\code{F}}{A_\code{F}}{S_\code{F}}$.
For the former, we first prove \code{\kwguarantee(x\%4 == 0)} succeeds since \code{x} is \code{40}; then for any given \code{r} from \code{F.f(x)},
we can assume \code{r\%4 == 1}, which implies \code{r\%2 == 1} thereby proving that both implementation and abspec print \code{42}.
For the latter, we can assume \code{x\%4 == 0} thereby proving that both implementation and abspec compute and print the same value \code{r}.
Then \code{\kwguarantee(r\%4 == 1)} succeeds since \code{r} is \code{(x/2 + 1)**2} and we assumed \code{x\%4 == 0}.
Finally, both return the same \code{r}.

Finally, we can see that the following hold:
\[
  \beh{\spcmod{S_\code{Main} \cup S_\code{F}}{A_\code{Main}}{S_\code{Main}} \link \spcmod{S_\code{Main} \cup S_\code{F}}{A_\code{F}}{S_\code{F}}} \subseteq \beh{A_\code{Main} \link A_\code{F}}
\]
which easily follows using the following lemma: for any proposition $P$ and any program with a hole $K[-]$,
\[
\beh{K[\kwguarantee(P); \kwassume(P)]} \subseteq \beh{K[\kwskip]}
\]
This lemma holds trivially if $P$ holds; otherwise it holds since the left hand side exhibits no behavior.


\subsection{Separation logic specifications in \CREMS{}}
\label{sec:overview:separation}

\begin{figure}[t]
$I_\Main$ :=
\begin{minipage}[t]{0.31\textwidth}
\begin{Verbatim}[commandchars=\\\{\},codes={\catcode`$=3},fontsize=\small]
[\kwmodule Main]
\kwdef main() $\equiv$
  \kwrepeat \numfire \{
    \kwvar r := \Cannon.\fire{}();
    \prt(r)
  \}
\end{Verbatim}
\end{minipage}%
\hfill%
$I_\Cannon$ :=
\begin{minipage}[t]{0.39\textwidth}
\begin{Verbatim}[commandchars=\\\{\},codes={\catcode`$=3},fontsize=\small]
[\kwmodule \Cannon{}]
\kwlocal \powder = 1
\kwdef \fire{}() $\equiv$
  \kwvar r := 1/\powder;
  \prt(r);
  \powder := \powder - 1;
  r
\end{Verbatim}
\end{minipage}%
\\[1mm]
\begin{minipage}[t]{0.41\textwidth}
$S_\Main$ :=
$\left\{
\begin{array}{@{}r@{}l@{}l@{}}
\hoare{\slball}{\code{Main.main}} {\TRUE}
\end{array}
\right\}$
\end{minipage}%
\hfill%
\begin{minipage}[t]{0.495\textwidth}
$S_\Cannon$ :=
$\left\{
\begin{array}{@{}r@{}l@{}l@{}}
\hoare{\slball} {\code{\Cannon.\fire}} {\Ret\valr. \valr = 1}
\end{array}
\right\}$
\end{minipage}%
\\[1mm]
\begin{minipage}[t]{0.4\textwidth}
$\mires_\Main$ := $\kwemp$
\end{minipage}%
\hfill%
\begin{minipage}[t]{0.495\textwidth}
$\mires_\Cannon$ := $\ready$
\end{minipage}%
\\[1mm]
$A_\Main$ :=
\begin{minipage}[t]{0.31\textwidth}
\begin{Verbatim}[commandchars=\\\{\},codes={\catcode`$=3},fontsize=\small]
[\kwmodule Main]
\kwdef main() $\equiv$
  \kwrepeat \numfire \{
    \kwvar r := \Cannon.\fire{}();
    \prt(\fbox{1})
  \}
\end{Verbatim}
\end{minipage}%
\hfill%
$A_\Cannon$ :=
\begin{minipage}[t]{0.39\textwidth}
\begin{Verbatim}[commandchars=\\\{\},codes={\catcode`$=3},fontsize=\small]
[\kwmodule \Cannon{}]
\fbox{\phantom{local n = 1}}
\kwdef \fire{}() $\equiv$
  \kwvar r := \fbox{1};
  \prt(\fbox{1});
  \fbox{\phantom{n := n - 1;}}
  r
\end{Verbatim}
\end{minipage}%
\\[1mm]
\begin{minipage}[t]{0.5\textwidth}
$\spcmod{S_\Main \cup S_\Cannon}{(A_\Main,\mires_\Main)}{S_\Main}$ :=\\
\hspace*{10pt}\begin{minipage}[t]{0.4\textwidth}  
\begin{Verbatim}[commandchars=\\\{\},codes={\catcode`$=3},fontsize=\small]
[\kwmodule Main]
\boilerplate{\kwlocal \kwmp :=} \hl{$\kwemp$}
\kwdef main() $\equiv$
  \boilerplate{\kwvar \kwfp := \kwemp;}
 \mydashbox{red}{14pc}{\ASSUME($\lambda$\res. \res == \ball)}
  \kwrepeat \numfire \{
   \mydashbox{blue}{13.25pc}{\GUARANTEE($\lambda$\res. \res == \ball)}
    \kwvar r := \Cannon.\fire{}();
   \mydashbox{red}{13.25pc}{\ASSUME($\lambda$\res. \res == \kwemp && r == 1)}
    \prt(1)
  \}
 \mydashbox{blue}{14pc}{\GUARANTEE($\lambda$\res. \res == \kwemp)}
\end{Verbatim}
\end{minipage}%
\end{minipage}%
\hfill%
\begin{minipage}[t]{0.5\textwidth}
$\spcmod{S_\Main \cup S_\Cannon}{(A_\Cannon,\mires_\Cannon)}{S_\Cannon}$ :=\\
\hspace*{10pt}\begin{minipage}[t]{0.4\textwidth}  
\begin{Verbatim}[commandchars=\\\{\},codes={\catcode`$=3},fontsize=\small]
[\kwmodule \Cannon{}]
\boilerplate{\kwlocal \kwmp :=} \hl{$\ready$}
\kwdef \fire{}() $\equiv$
  \boilerplate{\kwvar \kwfp := \kwemp;}
 \mydashbox{red}{14pc}{\ASSUME($\lambda$\res. \res == \ball)}
  \kwvar r := 1;
  \prt(1);
 \mydashbox{blue}{14pc}{\GUARANTEE($\lambda$\res. \res == \kwemp && r == 1)}
  r
\end{Verbatim}
\end{minipage}%
\end{minipage}%
\\[2mm]
\begin{minipage}[t]{0.5\textwidth}
\begin{Verbatim}[commandchars=\\\{\},codes={\catcode`$=3},fontsize=\small]
\color{red}\ASSUME(Cond) $\equiv$ \{
\color{red}  \kwvar \res := \kwtake(\kwcanonpcm);
\color{red}  \kwassume(Cond(\res));
\color{red}  \kwfp := \kwfp + \res;
\color{red}  \kwassume(\kwmp + \kwfp != \kwundef);
\color{red}\}
\end{Verbatim}
\end{minipage}%
\hfill%
\begin{minipage}[t]{0.5\textwidth}
\begin{Verbatim}[commandchars=\\\{\},codes={\catcode`$=3},fontsize=\small]
\color{blue}\GUARANTEE(Cond) $\equiv$ \{
\color{blue}  (\kwmp, \kwfp) := \kwfpu(\kwmp, \kwfp);
\color{blue}  \kwvar \res := \kwchoose(\kwcanonpcm);
\color{blue}  \kwguarantee(Cond(\res));
\color{blue}  \kwfp := \kwminus(\kwfp, \res);
\color{blue}\}
\end{Verbatim}
\end{minipage}%
\myhrule
\caption{Implementations, SL specifications, Initial resources, Abstractions,
  and Abspecs for \Main{} and \Cannon{}}
\label{fig:sep-logic}
\end{figure}



Now to see how we can express separation logic (SL) specifications in \CREMS,
consider the implementations $I_\Cannon$ and $I_\Main$  of the two modules \Main{} and \Cannon{}, shown in \Cref{fig:sep-logic}.
Here \code{\Main{}.\main{}()} invokes \code{\Cannon{}.\fire{}()} and prints the return value for a fixed number, \numfire{}, of times.
The \Cannon{} module consists of $(i)$ the \emph{module-local} variable \powder{} (\ie not accessible to other modules),
which is initially set to \code{1} and
$(ii)$ the function \fire{}, which sets the variable \code{r} to be \code{1/\powder{}}, prints \code{r}, decrements \powder{} by \code{1}, and returns \code{r}.
Note that \code{\Cannon{}.\fire{}()} can be safely invoked only once because it will crash due to division by zero at the second invocation;
therefore, if \numfire{} is \code{2}, the whole program crashes.

We briefly discuss how one can prove safety of the program when \numfire{} is \code{1} in SL.
First, the SL specification $S_\Cannon$ (given in \Cref{fig:sep-logic}) says that
if the resource $\ball$ is logically given, \code{\Cannon{}.\fire{}()} safely executes and returns \code{1}
but does not logically give the $\ball$ back (\ie $\ball$ is \emph{consumed}).
Here one can intuitively understand a \emph{resource} as something that is neither duplicable nor creatable out of nothing.
In this example, we can design the universe of resources---via a general mechanism based on PCMs (Partial Commutative Monoids)---in such a way that there can be at most one $\ball$,
relying on which we can then \emph{locally} prove the safety of \code{\Cannon{}.\fire{}()}
since it requires $\ball$ and thus can be invoked at most once.
Second, the SL specification $S_\Main$ says that given a $\ball$, \code{\Main{}.\main{}()} safely executes,
which is \emph{locally} provable relying on $S_\Cannon$ since \numfire{} is \code{1} and thus \code{\Cannon{}.\fire{}()} is invoked only once.
Indeed, if \numfire{} was \code{2}, one cannot verify \code{\Main{}.\main{}()} against $S_\Main$
since there is no $\ball$ left at the second invocation of \code{\Cannon{}.\fire{}()}.
Note that this kind of reasoning would not be possible in Hoare logic because
\code{\Cannon{}.\fire{}()} has no arguments, so that it would be hard to express any kind of precondition.

With this intuition, in \crems, assuming $S_\Main$ and $S_\Cannon$, we would like to \emph{locally} verify $I_\Main$ and $I_\Cannon$ against, \eg the abstractions $A_\Main$ and $A_\Cannon$ in \Cref{fig:sep-logic}, where the abstracted parts are \fbox{boxed}.
Note that without any assumption neither $I_\Main \lectx A_\Main$ nor $I_\Cannon \lectx A_\Cannon$ holds
since their context modules are arbitrary and hence, \eg \code{0} may be given for \code{r} in $\code{Main.main()}$, and also
\code{\Cannon{}.\fire{}()} may be invoked twice by the context in which case $I_\Cannon$ crashes but $A_\Cannon$ runs successfully.
As before, to solve this problem, we will \emph{internalize} the SL specifications inside the abstractions,
which are the abspecs given in \Cref{fig:sep-logic}:
\[
\spcmod{S_\Main \cup S_\Cannon}{(A_\Main,\mires_\Main)}{S_\Main} \text{ and }
\spcmod{S_\Main \cup S_\Cannon}{(A_\Cannon,\mires_\Cannon)}{S_\Cannon}
\]

To understand the abspecs, we first see how the set of resources $\kwcanonpcm$ is defined.
Concretely, $\kwcanonpcm$ consists of five elements $\kwundef, \kwemp, \ready, \fired, \ball$ with a commutative binary operator $+$ defined as follows:
\begin{itemize}
\item $\forall p \in \kwcanonpcm,~ \kwundef + p = \kwundef$
\item $\forall p \in \kwcanonpcm,~ \kwemp + p = p$
\item $\ready + \ball = \fired$
\item $\fired + \ready = \fired + \ball = \ready + \ready = \fired + \fired = \ball + \ball = \kwundef$
\end{itemize}
The intuition here is that $\kwundef$ represents undefinedness (\ie \emph{inconsistency}); $\kwemp$ the \emph{empty resource}, which is the identity for $+$;
$\ready$ the \emph{knowledge} that \Cannon{} is not yet fired;
$\fired$ that \Cannon{} is already fired;
and $\ball$ the \emph{capability} to invoke \code{\Cannon{}.\fire{}()}.
The definition of $+$ captures that only a subset of $\{\ready, \ball\}$ or $\{\fired\}$ is consistent; in particular,
there cannot be two (or more) \ball{s} since $\ball + \ball = \kwundef$.

Now we look at the abspecs.
The dotted boxes are generated from $S_\Main$ and $S_\Cannon$,
where \ASSUME{} and \GUARANTEE{} are the macros defined at the bottom of \Cref{fig:sep-logic}.
Also the \boilerplate{gray code} is boilerplate
and the \hl{highlighted parts} are modules' initial resources, which come from $\mires_\Main$ and $\mires_\Cannon$ of the abspecs.

Then we see how it works.
Each module has a module-local resource, \kwmp, initialized with its initial resource
and each function has a function local resource, \kwfp, initialized with \kwemp{}.
Then \code{\Cannon.\fire{}()} \emph{assumes} its precondition saying that a resource \ball{} is given at the beginning, and \emph{guarantees} its postcondition saying that no resource is returned and the return value \code{r} is \code{1} at the end.
\code{\Main.\main{}()} also \emph{assumes} its precondition saying that a resource \ball{} is given at the beginning; then \emph{guarantees} the precondition of \code{\Cannon.\fire{}()} before invoking it and \emph{assumes} the postcondition of \code{\Cannon.\fire{}()} after the invocation;
finally \emph{guarantees} its postcondition at the end.

Now we see the computational interpretation of \ASSUME{} and \GUARANTEE{}, whose macro definitions are given in \Cref{fig:sep-logic}.
First, \textcolor{red}{\code{\ASSUME(Cond)}} \emph{takes} a resource; assumes it satisfies \code{Cond};
adds it to \kwfp{}; and assumes \kwfp{} is consistent with \kwmp{}.
Here the \kwtake{} operation is a key technique in \crems and will be explained below.
For now, we simply consider as if \code{\kwtake(\kwcanonpcm)} magically takes the resource that is given by the caller or returned by the callee.
Second, \textcolor{blue}{\code{\GUARANTEE(Cond)}} $(i)$ nondeterministically updates \kwmp{} and \kwfp{} via \kwfpu{}, called \emph{frame-preserving update (FPU)}\footnote{
  The notion of frame-preserving update comes from modern separation logic such as \iris~\cite{irisgroundup}.
};
$(ii)$ nondeterministically \emph{chooses} a resource; $(iii)$ guarantees it satisfies \code{Cond}; and $(iv)$ \emph{subtract} it from \kwfp{}.
In $(i)$, we are allowed to update the module and function resources before jumping to another function,
which however is restricted to FPUs. The definition of \kwfpu{} is given as follows\footnote{
  Our formal definition of $\kwfpu$ is slightly more general as found in SLs (see \Cref{formal:embedding2} for details).
}:
\[
\begin{array}{@{}l@{}}
  \kwfpu(\kwmp,\kwfp) ~\defeq~ \kwchoose(\{ (\kwmp',\kwfp') \,|\, \forall \resframe \in \kwcanonpcm,~ \\
  \hspace*{1pc}\code{\resframe + $\kwmp$ + $\kwfp$ != \kwundef} \implies \code{\resframe + $\kwmp'$ + $\kwfp'$ != \kwundef} \})
\end{array}
\]
where $\kwchoose(X)$ nondeterministically picks an element from the given set $X$.
The intuition is that one can update $\kwmp$ and $\kwfp$ in such a way that consistency is preserved
under an arbitrary frame resource $\resframe$ (capturing possible resources remaining in other modules).
In $(ii)$, although a resource $\res$ is just nondeterministically chosen,
for now we simply consider as if it is magically passed to the callee or returned to the caller.
In $(iii)$, if the chosen $\res$ does not satisfy the condition $\code{Cond}$, it triggers no behavior,
which means that only that resource satisfying the condition can be chosen.
In $(iv)$, we performed the subtraction because we are passing it to another function, which will add it to its own function resource as we have seen above in \textcolor{red}{\code{\ASSUME(Cond)}}.
The definition of $\kwminus$ is given as follows:
\[
  \kwminus(\kwfp,\res) ~\defeq~ \kwchoose(\{ \kwfp' \,|\, \code{$\kwfp'$ + $\res$ == $\kwfp$} \})
\]
Note that when there is no such $\kwfp'$, it triggers no behavior because choosing from the empty set does so.

Now we see how the illusion of resource passing between functions works via \kwchoose{} and \kwtake{}.
First of all, note that it does not make sense to \emph{physically} pass any resource information to context modules or receive it from them
because context modules are completely arbitrary and thus may not understand such resource information (\eg those written in \imp).
Our key observation, however, is that by defining \kwtake{} as the dual operation to \kwchoose{}, we can logically make such an illusion.
Specifically, the nondeterministic choice \kwchoose{} and its dual \kwtake{}\footnote{
  The dual to (standard or demonic) nondeterminism is called \emph{angelic nondeterminism} in the literature~\cite{bodik2010programming, back2012refinement, tyrrell2006lattice, jeremie:lics}.
} are defined as follows for any set $X$ (See \Cref{sec:formal} for formal definitions).
\[
\begin{array}{rcl}
  \beh{\code{x := \kwchoose($X$); $K[\code{x}]$}} &\defeq& \bigcup_{x\in X} \beh{K[x]} \\
  \beh{\code{x := \kwtake($X$); $K[\code{x}]$}} &\defeq& \bigcap_{x\in X} \beh{K[x]} \\
\end{array}
\]
Note that $\kwchoose(\emptyset)$ is NB and $\kwtake(\emptyset)$ is UB.
The intuition is that
when proving $\beh{I} \subseteq \beh{\code{\res := \kwchoose(\kwcanonpcm); $K[\res]$}}$,
it suffices to prove $\beh{I} \subseteq \beh{K[\res]}$ for \emph{some} resource $\res$,
which allows us to \emph{logically} (\ie in the proof) pick a particular resource to pass to the context.
On the other hand,
when proving $\beh{I} \subseteq \beh{\code{\res := \kwtake(\kwcanonpcm); $K[\res]$}}$,
we need to prove $\beh{I} \subseteq \beh{K[\res]}$ for \emph{every} resource $\res$,
which makes sense
because we do not know what resource will be given from the context, and thus we have to prove the refinement whatever resource is given.
Then we have the following theorem, which makes the illusion of passing a resource from \kwchoose{} to \kwtake{}.
\begin{equation}
\begin{array}{@{}rl@{}}
&\beh{\code{\res{} := \kwchoose(\kwcanonpcm); $K[\res]$; $\res'$ := \kwtake(\kwcanonpcm); $K'[\res']$}}\\
\subseteq
&\beh{\code{\res{} := \kwchoose(\kwcanonpcm); $K[\res]$; $K'[\res]$}}
\end{array}\tag{$\star$}
\end{equation}
This theorem holds simply by instantiating the \kwtake{} operation with the chosen resource $\res$.

With this, we can prove $I_\Main \lectx \spcmod{S_\Main \cup S_\Cannon}{(A_\Main,\mires_\Main)}{S_\Main}$ for ${\numfire = 1}$.
When $\Main.\main()$ is invoked,
by \textcolor{red}{\code{\ASSUME($\lambda$\res. \res{} == \ball)}}
for any \emph{taken} resource $\res$, we can assume \code{\res{} == \ball} and thus add $\ball$ to $\kwfp$ yielding $\ball$.
Then
for \textcolor{blue}{\code{\GUARANTEE($\lambda$\res{}. \res{} == \ball)}}
the abspec does not update \kwmp{}, \kwfp{}, \emph{chooses} $\ball$ for $\res$ to satisfy the precondition of $\Cannon.\fire()$,
and successfully subtracts $\ball$ from $\kwfp$ yielding $\kwemp$.
Then both implementation and abspec invoke $\Cannon.\fire()$ and receive the same return value \code{r}.
By \textcolor{red}{\code{\ASSUME($\lambda$\res{}. \res{} == \kwemp \&\& r == 1)}}
for any \emph{taken} resource $\res$, we can assume \code{\res{} == \kwemp \&\& r == 1}, and thus add $\kwemp$ to $\kwfp$ yielding $\kwemp$.
Since $\code{r} = 1$, both implementation and abspec print \code{1}.
Finally,
we can trivially satisfy \textcolor{blue}{\code{\GUARANTEE($\lambda$\res{}. \res{} == \kwemp)}}
by not updating the module and function resources and \emph{choosing} $\kwemp$ for $\res$.
It is important to note that for ${\numfire = 2}$ the above proof breaks down
since at the second iteration we do not have $\ball$ anymore and thus cannot satisfy
\textcolor{blue}{\code{\GUARANTEE($\lambda$\res{}. \res{} == \ball)}}.

Similarly, we can prove $I_\Cannon \lectx \spcmod{S_\Main \cup S_\Cannon}{(A_\Cannon,\mires_\Cannon)}{S_\Cannon}$.
This time we set the module-local relational invariant to be $(\powder = 1 \land \kwmp = \ready) \lor (\powder = 0 \land \kwmp = \fired)$, which initially holds.
Then when $\Cannon.\fire()$ is invoked,
by \textcolor{red}{\code{\ASSUME($\lambda$\res{}. \res{} == \ball)}}
for any \emph{taken} resource $\res$, we can assume \code{\res{} == \ball} and thus add $\ball$ to $\kwfp$ yielding $\ball$.
Now due to the relational invariant, we have two cases.
First, when $\powder = 0 \land \kwmp = \fired$, the refinement trivially holds since $\kwmp + \kwfp = \kwundef$ and thus the abspec triggers UB.
Second, when $\powder = 1 \land \kwmp = \ready$, both implementation and abspec assign \code{1} to \code{r} and print \code{1};
the implementation decrements $\powder$ by \code{1} yielding \code{0} while
for \textcolor{blue}{\code{\GUARANTEE($\lambda$\res{}. \res{} == \kwemp \&\& r == 1)}}
the abspec updates $(\kwmp, \kwfp)$ to $(\fired, \kwemp)$, which is frame-preserving, and
\emph{chooses} $\kwemp$ for $\res$ to satisfy the postcondition and successfully subtract $\kwemp$ from $\kwfp$ yielding $\kwemp$;
finally both return the same value $\code{1}$ and establish the relational invariant with $\powder = 0 \land \kwmp = \fired$.
It is important to note that invariants relating implementations and abspecs like $(\powder = 1 \land \kwmp = \ready) \lor (\powder = 0 \land \kwmp = \fired)$ above do not appear in the abspecs, but only as a part of simulation proofs.

Finally, we can compose the abspecs to erase the specification parts:
\[
\begin{array}{@{}rl@{}}
&\beh{\spcmod{S_\Main \cup S_\Cannon}{(A_\Main,\mires_\Main)}{S_\Main} \link \spcmod{S_\Main \cup S_\Cannon}{(A_\Cannon,\mires_\Cannon)}{S_\Cannon}}\\
\subseteq & \beh{A_\Main \link A_\Cannon}
\end{array}
\]
We can prove this theorem, called \emph{spec erasure theorem}, as follows.
To \emph{discharge} the initial \ASSUME{} of $\Main.\main()$ (\ie to replace it with \kwskip),
we just need to choose an initial resource $\mires_\main$ to $\Main.\main()$, which will be $\ball$ here, and show that
$(i)$ it is consistent with the initial module-local resources (\ie \code{$\mires_\main$ + $\mires_\Main$ + $\mires_\Cannon$ != \kwundef{}})
and $(ii)$ it satisfies the precondition of $\Main.\main()$ (\ie \code{$\mires_\main$ == \ball}).
Then each remaining \textcolor{red}{\code{\ASSUME($Cond$)}} with a predicate $Cond$ on $\kwcanonpcm$ is discharged by the immediately preceding \textcolor{blue}{\code{\GUARANTEE($Cond$)}} with the same predicate.
To prove this, we first show that consistency of the whole resources (\ie those stored in all $\kwmp$ and $\kwfp$) is invariant:
the consistency holds initially because we have shown it by $(i)$ above,
and is preserved by \textcolor{blue}{\code{\GUARANTEE($Cond$)}}; \textcolor{red}{\code{\ASSUME($Cond$)}}
because the frame-preserving update via $\kwfpu$ preserves it by definition
and, by $(\star)$ above, the subtracted resource $\res$ in \GUARANTEE{} is immediately added back in \ASSUME{}.
Then we can complete the proof by discharging all the assumptions in \ASSUME{}, again by $(\star)$:
\textcolor{red}{\code{\kwassume($Cond$(\res))}} is discharged by \textcolor{blue}{\code{\kwguarantee($Cond$(\res))}},
and \textcolor{red}{\code{\kwassume(\kwmp{} + \kwfp{} != \kwundef)}} by the above invariant (\ie consistency of the whole resources).
Note that the spec erasure theorem holds in general among any compatible abspecs:
for example, even when ${\numfire = 2}$, the above erasure holds.

To conclude, the most important idea in \crems is
to give an illusion of passing logical information
via \kwchoose{} and \kwtake{} in an operational way.
In the next section, we will see how this powerful mechanism can be used
to model various logical features seamlessly.





\section{Advanced Features and Examples of \CREMS{}}
\label{sec:advanced}

In this section, we will demonstrate a general version of \CREMS with five advanced features:
$(i)$ dealing with unknown contexts, $(ii)$ proving and exploiting purity (\ie absence of side effects),
$(iii)$ decomposing and reusing verification tasks via gradual abstraction, $(iv)$ abstracting function arguments and return values, and $(v)$ reasoning about function pointers.
We will introduce them by walking through various examples.

\subsection{Dealing with unknown contexts}
\label{sec:advanced-unknown}

\begin{figure}[t]
$I_\Mem$ :=
\begin{minipage}[t]{0.42\textwidth}
\begin{Verbatim}[commandchars=\\\{\},codes={\catcode`$=3},fontsize=\small]
[\kwmodule Mem]
\kwlocal mem := ...
\kwdef alloc([n: \vint{}]\unwrapU{}) $\equiv$ ...
\kwdef free([p: \vptr{}]\unwrapU{}) $\equiv$ ...
\kwdef load([p: \vptr{}]\unwrapU{}) $\equiv$ ...
\kwdef store([p: \vptr{}, v: \vval{}]\unwrapU{}) $\equiv$ ...
\end{Verbatim}
\end{minipage}%
\hfill%
$A_\Mem$ :=
\begin{minipage}[t]{0.42\textwidth}
\begin{Verbatim}[commandchars=\\\{\},codes={\catcode`$=3},fontsize=\small]
[\kwmodule Mem]
\kwlocal mem := ...
\kwdef alloc =
 \kwfriend(\_) $\equiv$ \kwpure
 \kwcontext([n: \vint{}]\unwrapU{}) $\equiv$ ...
... 
\end{Verbatim}
\end{minipage}
\\[2mm]
\begin{minipage}[t]{0.5\textwidth}
$\mires_\Mem$ := $\authfull\munit \in \authm{\,(\vptr \rightarrow \exm{\,(\vval)})} \subseteq \Sigma$
\end{minipage}%
\hfill%
\begin{minipage}[t]{0.5\textwidth}
\end{minipage}
\\[1mm]
\begin{minipage}[t]{\textwidth}
\small$\begin{array}{@{}l@{}l@{\;}l@{\;}l@{}}
S_\Mem := \{&
\code{\Mem.alloc}{:}
&    
\forall n:\vint.
&
\curlybracket{\lambda\,\argp\,\_\,\argo.\; \rplift{\argo = \opure\;\_ \lland x = \upcast{[n]} \lland n \ge 0}}
\\&
&&
\curlybracket{\lambda\,\retp\,\_.\; \exists p:\vptr, \lst: \vlist\;\vval.\; \rplift{\retp = \upcast{p} \lland \lstlength(\lst) = n} * {(p \mapsto \lst)}},
\\&
\code{\Mem.free}{:}
&
\forall \_:\vunit.
&
\curlybracket{\lambda\,\argp\,\_\,\argo.\; \exists p:\vptr,v:\vval.\; \rplift{\argo = \opure\;\_ \lland \argp = \upcast{[p]}} * (p \mapsto [v])}
\\&
&&
\curlybracket{\lambda\,\retp\,\_.\; \rplift{\retp \in \upcast{\vval}}},
\\&
\code{\Mem.load}{:}
&
\forall (p,v): \vptr\!\times\!\vval.
&
\curlybracket{\lambda\,\argp\,\_\,\argo.\; \rplift{\argo = \opure\;\_ \lland \argp = \upcast{[p]}} * (p \mapsto [v])}
\\&
&&
\curlybracket{\lambda\,\retp\,\_.\; \rplift{ \retp = \upcast{v}} * (p \mapsto [v])},
\\&
\code{\Mem.store}{:}
&
\forall (p,v): \vptr\!\times\!\vval.
&
\curlybracket{\lambda\,\argp\,\_\,\argo.\; \rplift{\argo = \opure\;\_ \lland \argp = \upcast{[p, v]}} * \exists v':\vval.\; p \mapsto [v']}
\\&
&&
\curlybracket{\lambda\,\retp\,\_.\; \rplift{\retp \in \upcast{\vval}} * (p \mapsto [v])}
\hfill\}
\end{array}$
\end{minipage}
\myhrule
\caption{An implementation and its abspec for the module \Mem{}.}
\label{fig:mem}
\end{figure}



Unlike in the previous section where we composed abspecs for the \emph{whole} modules,
here we will see how to compose those for only selected modules, called \emph{friends},
while still allowing them to interact with arbitrary context modules, called \emph{contexts}
(\ie establishing contextual refinement as a result).

For this, our first observation is that there is a specification that every module satisfies.
Specifically, one can easily see the following holds: for any module $C$ in \ems,
\[
C \lectx \spcmod{\Snone}{(C,\mbox{$\munit$})}{\Snone}
\]
where $\munit$ is the identity for a PCM in consideration
and $\Snone$ has $\snone = \hoare{\TRUE}{}{\TRUE}$ for every function $f$ in the scope
(\ie the first $\Snone$ above specifies all the functions invoked by $C$ while the second $\Snone$ all the functions defined by $C$).
The essential reason why \emph{every} module can satisfy the above CR is because specifications in \crems do not require \emph{safety}, unlike in SL.

Although we can give specifications to arbitrary contexts as above, we still have to address
the problem that the specifications for contexts are not compatible with those for friends.
Specifically,
the contexts assume $\Snone$ for each module $F$ among the friends while its specification $S_F$ may not be $\Snone$.
We solve this problem by allowing abspecs to provide two different behaviors:
those satisfying intended specifications when invoked by the friends;
and those satisfying $\Snone$ when invoked by the contexts.
To enable this, we add a special mechanism to \ems that allows a callee to get the caller's module name,
so that the callee can behave differently depending on who's the caller.

As an example, consider the module $\Mem$ given in \Cref{fig:mem}.
The implementation $I_\Mem$ is directly written in \ems, which provides a CompCert-like memory model for the \imp language.
Here, to see what, \eg \code{[p: \vptr{}, v: \vval{}]\unwrapU{}} in \code{store} means,
note that in \ems every function takes a value of type \anyty{}, which can be seen as the set of all mathematical values,
and also returns an \anyty{} value, while the \imp language supports values of type \vval{}\footnote{
  To support a compilation to CompCert, \vval{} also contains the special value \kw{undef}.
}
consisting of 64-bit integers (\vint) and function and memory pointers (\vptr).
Therefore, when defining the semantics of \imp in \ems,
we \emph{pack} arguments to a function into a single value of type $\vlist\;\vval$ and \emph{upcast} it into $\anyty{}$,
and conversely \emph{downcast} and \emph{unpack} a given \anyty{} argument to a list of values of expected types,
where we trigger \kwub{} if the downcast or the unpacking fails.
The notation \code{[p: \vptr{}, v: \vval{}]\unwrapU{}} denotes such downcast and unpacking of an \anyty{} argument into a list of two values of types \vptr{} and \vval{}.
Now we see what $I_\Mem$ does.
\code{alloc(n)} allocates a memory block consisting of \code{n} cells, each of which can store a value of type \vval{},
and returns the pointer pointing to the beginning of the block;
\code{free(p)} deallocates the cell pointed to by \code{p};
\code{load(p)} reads a value from the cell pointed to by \code{p} and returns it;
and \code{store(p, v)} stores the value \code{v} in the cell pointed to by \code{p}.

Now we see how the abspec for $\Mem$, given in \Cref{fig:mem}, is defined,
where each function consists of two definitions, \kwfriend{} and \kwcontext{}.
Concretely, the \kwfriend{} definitions are all \kwpure{}, which can be simply understood as \kwskip{} for now (see \Cref{sec:advanced-purity} for details),
while the \kwcontext{} definitions are identical to their implementations.
The intention is that the former defines its behaviors when invoked by friends and the latter by contexts.
Although we do not know which modules will be friends yet,
we can still locally verify the following CR for any specification $S$
with $\mires_\Mem$, $S_\Mem$ given in \Cref{fig:mem},
\[
I_\Mem \lectx \spcmod{S}{(A_\Mem, \mires_\Mem)}{S_\Mem}
\]
Here the \ems semantics of each function $f$ in $\spcmod{S}{(A_\Mem, \mires_\Mem)}{S_\Mem}$
is given by \emph{intersecting} the semantics of \kwfriend{} and that of \kwcontext{}.
Specifically, we can intersect them by:
\[
\code{\kwvar{} b = \kwtake(\vbool); \kwif{} (b) \kwthen{} $C_\kwfriend$ \kwelse{} $C_\kwcontext$}
\]
where the semantics $C_\kwfriend$ for $f$ is generated from the \kwfriend{} definition of $f$ in $A_\Mem$
together with its specification $S \rtimes S_\Mem$ (and $\mires_\Mem$)
in the way we have seen in the previous section;
similarly for $C_\kwcontext$ but with the \kwcontext{} definition and $S \rtimes \Snone$.
Such intersection makes sense because it is essential to establish refinement between $I_\Mem$ and $A_\Mem$ for both friends and contexts.
It is important to note that the specification $S \rtimes \Snone$ for $C_\kwcontext$
means that when we prove $C_\kwcontext$ satisfies $\Snone$,
we can still rely on the intended assumptions $S$ about friends when it invokes their functions.
We will see such examples in \Cref{sec:advanced-purity} and \Cref{sec:advanced-args}.

For such abspecs with \kwfriend{} and \kwcontext{}, we have a general spec erasure theorem yielding contextual refinement (\ie under arbitrary contexts).
\begin{theorem}\label{thm:spec-erasure}
Given a global PCM $\Sigma$ (including all PCMs of interest) and abspecs $\spcmod{S}{(A_i, \mires_i)}{S_i}$ w.r.t. $\Sigma$ for $i\in\{1,\ldots,n\}$ with any $S \sqsupseteq S_1 \cup \ldots \cup S_n$,
suppose that their module names (\ie friends) are $N = \setofz{\name_1, \ldots, \name_n}$, and
for any argument value $v$ to $\Main.\main$,
there is an initial resource $\mires$ to $\Main.\main$ such that
\code{$\mires$ + $\mires_1$ + $\ldots$ + $\mires_n$ != \kwundef} and,
if $\Main.\main$ is among the friends, $(v,\mires)$ satisfies its precondition.
Then we have the following:
\[
\spcmod{S}{(A_1, \mires_1)}{S_1} \link \ldots \link \spcmod{S}{(A_n, \mires_n)}{S_n}
\lectx
\toabs{A_1}{N}  \link \ldots \link \toabs{A_n}{N}
\]
\end{theorem}
\noindent
Here we can understand $S \sqsupseteq S'$ as $S \supseteq S'$ though it has a slightly more general definition (see~\Cref{sec:formal} for details).
Also the semantics of a function $f$ in $\toabs{A_i}{N}$ is defined
by combining its \kwfriend{} semantics $C_\kwfriend$ and its \kwcontext{} semantics $C_\kwcontext$ in $A_i$ as follows:
\[
\code{\kwif{} ($\kwgetcaller{}() \in N$) \kwthen{} $C_\kwfriend$ \kwelse{} $C_\kwcontext$}
\]
where $\kwgetcaller{}()$ is supported by \ems and returns the module name of the caller.
Also we henceforth call such $A_i$ \emph{pre-abstraction} and such $\toabs{A_i}{N}$ \emph{abstraction}.
Note that since the theorem establishes \emph{contextual refinement},
the contexts are \emph{completely unrestricted} (\eg they may be unsafe, make system calls, or make mutually recursive calls to the friends).

Now we discuss the details of $S_\Mem$.
Here we ignore the measure parameter $\argo$ (used to prove termination) and $\_$ in $S_\Mem$,
which will be discussed in \Cref{sec:advanced-purity} and \Cref{sec:advanced-args}.
First, $\upcast{}$ is the \emph{upcast} operator (\ie $\upcast{x}$ is a value of type \anyty{} for $x$ of any type $X$)
and $\upcast{\vval}$ is $\setofz{\upcast{v} \,|\, v \in \vval}$.
The parameters $\argp$ and $\retp$ are bound to argument and return values.
Then $S_\Mem$ is a standard specification for such memory operations that one would write in modern separation logics such as \iris~\cite{irisgroundup}.
Specifically, the pre and post conditions define predicates on resources (\ie $\Sigma \rightarrow \mathbf{Prop}$, which we call $\rProp$)
when values for the quantifiers and argument are given;
and the separating conjunction~$*$, magic wand $\wand$, lifting $\rplift{}$ of $\mathbf{Prop}$ to $\rProp$,
and existential and universal quantifiers for $\rProp$ are defined in the standard way.
Also, the PCM $\authm{\,(\vptr \rightarrow \exm{\,(\vval)})}$ is a standard authoritative PCM
and the points-to predicate \mbox{\code{p $\mapsto$ $\lst$}}, capturing that the pointer $p$ points to the beginning of consecutive cells that contain the values in~$\lst$, is derived in the standard way satisfying the following law:
\[
  {p \mapsto v :: \lst} \iff (p \mapsto [v]) * (p+8 \mapsto \lst)
\]
Note that in \crems, universal quantifiers such as $\forall (p,v): \vptr\times\vval$ above are also modeled via \kwchoose{} and \kwtake{}.
The reason is because values for those quantifiers are essentially determined by the callers, and
therefore the caller \emph{chooses} a value for the quantifier and the callee \emph{takes} the value as we have seen in the previous section.

Now we briefly discuss how to verify $I_\Mem$: for any specification $S$,
\begin{equation}
I_\Mem \lectx \spcmod{S}{(A_\Mem, \mires_\Mem)}{S_\Mem}\label{cr:mem}
\end{equation}
As usual, we first set up a module-local relational invariant and prove refinement between $I_\Mem$ and $\spcmod{S}{(A_\Mem, \mires_\Mem)}{S_\Mem}$,
which is split into two cases because the abspec is defined as the intersection of \kwfriend{} and \kwcontext{}:
proving $(i)$ that $I_\Mem$ refines the \kwfriend{} definitions of $A_\Mem$ under $S \rtimes S_\Mem$
and $(ii)$ that $I_\Mem$ refines the \kwcontext{} definitions of $A_\Mem$ under $S \rtimes \Snone$,
where both proofs involve preservation of the (common) relational invariant,
which essentially captures that the \kwfriend{} and \kwcontext{} definitions work in harmony (\ie they do not interfere each other's reasoning in any interleaved invocations of the two definitions).
Specifically, the invariant says that the blocks allocated in the implementation are split into two groups such that
in the abspec,
one of the groups is allocated at the same addresses\footnote{This is possible because our allocator is nondeterministic.} (performed by \kwcontext{})
and the other resides in the module-local resource (but not in the memory) in terms of the points-to predicate (performed by \kwfriend{}).
The reason why this invariant is preserved even when the \kwcontext{} definitions are invoked with arbitrary pointers
is essentially because when a context tries to access the blocks allocated by friends by forging their addresses,
the invariant guarantees that in the abspec no blocks are allocated at those addresses so that such accesses always trigger \kwub{},
which immediately completes the refinement.
Except the preservation of the invariant,
the proof of $(ii)$ is straightforward because it establishes refinement between identical definitions;
and that of $(i)$ essentially amounts to a standard SL proof for those specifications together with a termination proof,
which will be discussed in the following section.

\subsection{Proving and exploiting purity}
\label{sec:advanced-purity}

\begin{figure}[t]
$I_\Stack$ :=
\begin{minipage}[t]{0.33\textwidth}
\begin{Verbatim}[commandchars=\\\{\},codes={\catcode`$=3},fontsize=\small]
[\kwmodule Stack]
\kwdef new([]\unwrapU{}) $\equiv$
  \kwvar stk := Mem.alloc(1);
  Mem.store(stk, NULL);
  stk
\kwdef push([stk: \vval{}, v: \vval{}]\unwrapU{}) $\equiv$
  \kwvar node := Mem.alloc(2);
  \kwvar hd := Mem.load(stk);
  Mem.store(node, v);
  Mem.store(node+8, hd);
  Mem.store(stk,node)
\kwdef pop([stk: \vval{}]\unwrapU{}) $\equiv$
  \kwvar hd := Mem.load(stk);
  \kwif hd == NULL \kwthen 0
  \kwelse
    \kwvar v := Mem.load(hd);
    \kwvar next := Mem.load(hd+8);
    Mem.store(stk, next);
    Mem.free(hd); Mem.free(hd+8);
    v
\end{Verbatim}
\end{minipage}%
\hfill%
$A^1_\Stack$ :=
\begin{minipage}[t]{0.47\textwidth}
\begin{Verbatim}[commandchars=\\\{\},codes={\catcode`$=3},fontsize=\small]
[\kwmodule Stack]
\kwlocal pool := ($\lambda$_. \none)
  : \vptr {$\rightarrow$} \option (\vlist \vval{})
\kwdef new =
 \kwfriend,\kwcontext([]\unwrapU{}) $\equiv$
  \kwvar handle := \kwchoose(\vptr{});
  \kwguarantee(pool handle == \none);
  pool := pool[handle := \some []];
  handle
\kwdef push =
 \kwfriend,\kwcontext([handle: \vptr{}, v: \vval{}]\unwrapU{}) $\equiv$  
  \kwvar stk := unopt\unwrapU{}(pool handle);
  pool := pool[handle := \some (x::stk)]
\kwdef pop =
 \kwfriend,\kwcontext([handle: \vptr{}]\unwrapU{}) $\equiv$
  \kwvar stk := unopt\unwrapU{}(pool handle);
  \kwmatch stk \kwwith | [] => 0
  | v :: stk' =>
    pool := pool[handle := \some stk'];
    v  \kwend
\end{Verbatim}
\end{minipage}
\\[2mm]
$S^1_\Stack$ := $\setofz{\  \code{\Stack.new}:\snone,\  \code{\Stack.push}:\snone,\ \code{\Stack.pop}:\snone\ }$
\hspace*{5pc}$\mires^1_\Stack$ := $\munit$\hfill\mbox{}\\[1mm]
where $\snone = \forall \_:\vunit.\;\hoare{\lambda\,\argp\,\argv\,\argo.\; \rplift{\argo = \otop \lland \argp = \argv}}{}{\lambda\,\retp\,\retv.\; \rplift{\retp = \retv}}$\hfill\mbox{}
\myhrule
\caption{An implementation and its first abspec for the module \Stack{}.}
\label{fig:stack1}
\end{figure}



Now we discuss how to express, prove and exploit \emph{purity} of an abspec of a function.
Note that it is possible that even though an implementation has impurity,
if the impurity only changes the module's local state,
its abspec can be made pure by migrating the impurity to its specification (\ie pre and post conditions).
Indeed this is the case for the module $\Mem$ and we will see how we can do it in this section.

To see this clearly, we need another example, 
which is the module $\Stack$ given in \Cref{fig:stack1} and implemented using the module $\Mem$.
Concretely, $I_\Stack$ presents the \ems semantics of an \imp program\footnote{
  The downcast to and upcast from $\vval$ (with \kwub{} in case of failure) around a function call are omitted for syntactic clarity.
}
that implements stacks using linked lists,
where \code{new()} creates a new stack; \code{push(stk,v)} pushes the value \code{v} into the stack \code{stk};
and \code{pop(stk)} pops a value from \code{stk} and returns it if \code{stk} is nonempty; otherwise returns \code{\exitval}.

Then $A^1_\Stack$ presents an abstract version of the functions,
which \emph{module-locally} manage a pool of mathematical lists using $\vptr$ values as their handles
(defined as \code{\vptr{} {$\rightarrow$} \option{} (\vlist{} \vval)}).
Concretely, \code{new()} nondeterministically chooses an \emph{unused} handle (via \kwchoose{} and \kwguarantee{}),
registers the empty list with the handle in the pool and returns the handle;
\code{push(handle,v)} gets the list with \code{handle} from the pool (if fails, trigger \kwub{}) by \code{unopt\unwrapU{}(pool handle)}
and then updates the list by adding the value~\code{v};
and \code{pop(handle)} also gets the list with \code{handle} by \code{unopt\unwrapU{}(pool handle)}, and
if it is empty, returns \code{\exitval}; otherwise returns the head of the list after eliminating it from the list.
Note that in the pre-abstraction, the \kwfriend{} and \kwcontext{} definitions coincide and so do their specifications (\ie both are $\snone$,
the details of which will be explained in \Cref{sec:advanced-args}).

First of all, we start by noting the problem that it actually does not make sense to prove CR between the implementation and the abspec for $\Stack$
because there are several calls to $\Mem$ in $I_\Stack$ while there are no such calls in $A^1_\Stack$.
Even though $\Mem$ is specified in the abspec, when locally establishing CR for \code{Stack}, $\Mem$ is still arbitrary and thus can make arbitrary side effects such as fatal errors or making system calls.
Therefore completely abstracting away those calls in the pre-abstraction would not allow us to establish the desired CR.

Our approach to address this problem is three-fold:
$(i)$ giving a way of enforcing and specifying \emph{purity} (\ie the absence of side effects),
$(ii)$ giving an illusion of abstracting away those calls to functions specified as pure, simply called \emph{pure calls}, 
instead of actually eliminating them in the abspec,
and then
$(iii)$ truly eliminating them in the abstractions after applying the spec erasure theorem.

We discuss $(ii)$ first.
To give an illusion of eliminating pure calls in local reasoning as seen in $A^1_\Stack$, we introduce a mechanism, called \emph{implicit pure calls (IPC)}.
The idea is to implicitly introduce all possible pure calls (according to the given specification)
since they will be eliminated by the spec erasure theorem anyway.
Specifically, an IPC nondeterministically makes an unbounded but finite number of all possible pure calls with all possible (physical and logical) arguments
and we implicitly insert an IPC at each line (\eg the semicolon in $A^1_\Stack$ is interpreted as making an IPC).
Therefore without explicitly writing down specific pure calls,
we can freely rely on the specification of any pure function during the proof.

Now we discuss $(i)$: how to enforce and specify purity in \crems.
The notion of purity of a function $f$, saying that $f$ does not produce any side effects, can be mostly enforced by defining the pre-abstraction of $f$ as an IPC.
Since IPC only allows invoking pure functions, verifying against IPC essentially amounts to proving the absence of side effects except for non-termination.
Therefore, the remaining questions are how to enforce termination, how to specify purity and how to make a pure call.

We can answer all the question by adding the \emph{measure parameter} $\argo$ to preconditions, which can be passed from a caller to the callee via \kwchoose{} and \kwtake{}.
First, we define the type of measures to be $\option\;\ordinal$ with $\ordinal$ the set of ordinals\footnote{We developed our own Coq library for ordinals, which will be published elsewhere.} with a well-founded order $<$
and define a relation $\sqsubset$ between them by the following two cases:
\[
\opure\;o \sqsubset \opure\;o' \text{ with } o < o' \in \ordinal  \qquad\qquad  \argo \sqsubset \none \text{ with } \argo \in \option\;\ordinal
\]
Second, invoking a function with a measure $\opure\;\_$ is considered as a pure call and
the \ems semantics of a function in an abspec, when invoked with $\opure\;\_$, is defined to be an IPC.
Third, to enforce termination, when a function $f$ is invoked with a measure $\argo$,
for each function call made inside the invocation
we add the \kwguarantee{} that it should pass (\ie \kwchoose{}) a measure $\argo'$ with $\argo' \sqsubset \argo$.
Then it is guaranteed that a pure call (\ie with $\opure\;o$) can only make pure calls as sub calls (thereby producing no side effects other than non-termination) and should terminate
because any measure $\argo \sqsubset \opure\;o$ should be $\opure\;o'$ with $o' < o$.
Note that an impure call (\ie with the measure $\none$) can still make any calls
because we have $\argo \sqsubset \none$ for any measure $\argo$ including $\none$.

Then, we achieve $(iii)$: the spec erasure theorem can soundly eliminate all IPCs in the resulting abstractions.
Also, note that the \kwfriend{} definition of $f$ in an abspec only describe the impure behavior of $f$ since its pure behavior is defined to be an IPC.

With this in mind, we revisit the module $\Mem$.
First, all the specifications have $\argo = \opure\; \_$ in their preconditions, which implies that only pure calls to those functions can be made.
Second, since the functions in $\Mem$ do not allow any impure calls to them, their \kwfriend{} definitions are never executed,
which we can guarantee by defining them as $\kwnb$ (\ie \kwpure = \kwnb).
Then, in the abstraction $\toabs{A_\Mem}{N}$ for $\Mem$ after applying the spec erasure theorem with friends $N$,
we have the functions that trigger \kwnb{} if invoked by the friends, and do the original jobs otherwise.
This implies that the friends are guaranteed not to invoke any memory operations
(\ie all the calls to $\Mem$ by the friends are eliminated in their abstractions).
Note that using the fact that \kwnb{} contextually refines any possible definitions,
we can also revert the abstraction $\toabs{A_\Mem}{N}$ back to $I_\Mem$ (\ie $\toabs{A_\Mem}{N} \lectx I_\Mem$).
Also note that as discussed above, the implementation $I_\Mem$ can be seen as impure because, \eg calls to \kwstore{} have impact on subsequent calls to \kwload{},
while its pre-abstraction, IPC, for friends are indeed pure and thus can be eliminated.

Finally, we can verify $I_\Stack$: for any $S \sqsupseteq S_\Mem$, 
\begin{equation}
I_\Stack \lectx \spcmod{S}{(A^1_\Stack,\mires^1_\Stack)}{S^1_\Stack} \label{cr:stack1}
\end{equation}
For this, we set up the module-local invariant saying that
the lists in the pool are matched with the linked lists stored in the module-local resource of $\Stack$ in terms of the points-to predicate,
which are obtained via IPCs to the memory functions with $S_\Mem$.
Then, one can easily establish a simulation proof preserving the invariant.

\subsection{Decomposing and reusing verification tasks via gradual abstraction}
\label{sec:advanced-gradual}

\begin{figure}[t]
$A^2_\Stack$ := 
\begin{minipage}[t]{0.37\textwidth}
\begin{Verbatim}[commandchars=\\\{\},codes={\catcode`$=3},fontsize=\small]
[\kwmodule Stack]
\kwlocal pool := ($\lambda$_. \none)
  : \vptr {$\rightarrow$} \option (\vlist \vval)
\end{Verbatim}
\end{minipage}%
\begin{minipage}[t]{0.27\textwidth}
\begin{Verbatim}[commandchars=\\\{\},codes={\catcode`$=3},fontsize=\small]
\kwdef new =
 \kwfriend(\_) $\equiv$ \kwnb
 \kwcontext([]\unwrapU{}) $\equiv$ ...
\end{Verbatim}
\end{minipage}%
\begin{minipage}[t]{0.1\textwidth}
\begin{Verbatim}[commandchars=\\\{\},codes={\catcode`$=3},fontsize=\small]
...


\end{Verbatim}
\end{minipage}%
\hfill\mbox{}
\\[1mm]
$\mires^\twoA_\Stack$ := $\authfull\munit \in \authm{\,(\vptr \rightarrow \exm{\,(\code{\vlist \vval})})} \subseteq \Sigma$\hfill\mbox{}
\\[1mm]
\begin{minipage}[t]{\textwidth}
\small$\begin{array}{@{}l@{}l@{\;}l@{\;}l@{}}
S^\twoA_\Stack := \{
&    
\code{Stack.new}{:}
&    
\forall \_:\vunit.
&
\curlybracket{\lambda\,\argp\,\_\,\argo.\; \rplift{\argo = \opure\;\_ \lland \argp = \upcast{[]}}}
\\&
&&
\curlybracket{\lambda\,\retp\,\_.\; \exists \handle:\vptr.\; \rplift{\retp = \upcast{\handle}} * \isstack{\handle}{[]}},
\\&
\code{Stack.push}{:}
&
\forall (\handle,v,\lst):\vptr \!\times\! \vval \!\times\! \vlist\;\vval.
&
\curlybracket{\lambda\,\argp\,\_\,\argo.\; \rplift{\argo = \opure\;\_ \lland \argp = \upcast{[\handle, v]}} * \isstack{\handle}{\lst}}
\\&
&&
\curlybracket{\lambda\,\retp\,\_.\; \rplift{\retp \in \upcast{\vval}} * \isstack{\handle}{(v :: \lst)}},
\\&
\code{Stack.pop}{:}
&
\forall (\handle,\lst):\vptr \times \vlist\;\vval.
&
\curlybracket{\lambda\,\argp\,\_\,\argo.\; \rplift{\argo = \opure\;\_ \lland \argp = \upcast{[\handle]}} * \isstack{\handle}{\lst}}
\\&
&&
\curlybracket{\lambda\,\retp\,\_.\; \rplift{ \retp = \upcast{\lsthead(\lst,0)}} * \isstack{\handle}{\lsttail(\lst)}}
\hfill\}
\end{array}$
\end{minipage}
\\[1mm]
$\mires^\twoB_\Stack$ := $\authfull\munit \in \authm{\,(\vptr \rightarrow \optionm{\,(\agm{\,(\mathcal{P}(\vval))})})} \subseteq \Sigma$\hfill\mbox{}
\\[1mm]
\begin{minipage}[t]{\textwidth}
\small$\begin{array}{@{}l@{}l@{\;}l@{\;}l@{}}
S^\twoB_\Stack := \{
&  
\code{Stack.new}{:}
&
\forall P:\mathcal{P}(\vval).
&
\curlybracket{\lambda\,\argp\,\_\,\argo.\; \rplift{\argo = \opure\;\_ \lland \argp = \upcast{[]}}}
\\&
&&
\curlybracket{\lambda\,\retp\,\_.\; \exists \handle:\vptr.\; \rplift{\retp = \upcast{\handle}} * \isbag{\handle}{P}},
\\&
\code{Stack.push}{:}
&
\forall (\handle,v,P): \vptr \times \vval \ \ \ .
&
\curlybracket{\lambda\,\argp\,\_\,\argo.\; \rplift{\argo = \opure\;\_ \lland \argp = \upcast{[\handle, v]} \lland v\in P} * \isbag{\handle}{P}}
\\&
&\hspace*{3.2pc}{} \times \mathcal{P}(\vval)&
\curlybracket{\lambda\,\retp\,\_.\; \rplift{\retp \in \upcast{\vval}} * \isbag{\handle}{P}},
\\&
\code{Stack.pop}{:}
&
\forall (\handle,P): \vptr \times \mathcal{P}(\vval).
&
\curlybracket{\lambda\,\argp\,\_\,\argo.\; \rplift{\argo = \opure\;\_ \lland \argp = \upcast{[\handle]}} * \isbag{\handle}{P}}
\\&
&&
\curlybracket{\lambda\,\retp\,\_.\; \exists v:\vval.\; \rplift {\retp = \upcast{v} \lland (v = 0 \lor v \in P)} * \isbag{\handle}{P}}
\hfill\}
\end{array}$
\end{minipage}
\myhrule
\caption{Two abspecs on top of the first abstraction for the module \Stack{}}
\label{fig:stack2}
\end{figure}



Although the abstraction $\toabs{A^1_\Stack}{\setofz{\Mem,\Stack}}$, obtained by applying the spec erasure theorem for the friends $\Mem$ and $\Stack$,
can be directly used by other modules, 
it would be better to provide useful logical specifications for $\Stack$ like those we provided for $\Mem$.

\Cref{fig:stack2} shows two such specifications $S^\twoA_\Stack$ and $S^\twoB_\Stack$ for $\Stack$.
Then we can separately verify the previous abstraction against the two specifications together with the \emph{pure} pre-abstraction $A^2_\Stack$.
Specifically, we prove the following: for any specification $S$,\\
\begin{equation}
  \toabs{A^1_\Stack}{\setofz{\Mem,\Stack}} \ \ \lectx\ \  \spcmod{S}{(A^2_\Stack,\mires^\twoA_\Stack)}{S^\twoA_\Stack} \label{cr:stack2A}
\end{equation}
\begin{equation}
  \toabs{A^1_\Stack}{\setofz{\Mem,\Stack}} \ \ \lectx\ \  \spcmod{S}{(A^2_\Stack,\mires^\twoB_\Stack)}{S^\twoB_\Stack} \label{cr:stack2B}
\end{equation}
Here both $S^\twoA_\Stack$ and $S^\twoB_\Stack$ only allow pure calls to the functions by requiring $\argo = \opure\; \_$
and the pre-abstraction $A^2_\Stack$ is the same as $A^1_\Stack$ for \kwcontext{}, and \kwnb{} for \kwfriend{},
as we have done for $\Mem$.
The two specifications provide different benefits to the client:
the former precisely tracks the contents of a stack via the predicate $\isstack{\handle}{\lst}$ saying that
the stack with handle $\handle$ in the pool coincides with $\lst$,
while the latter maintains a certain property for a stack via the predicate $\isbag{\handle}{P}$
saying that all the elements in the stack with handle $\handle$ satisfy the property $P$
and furthermore allows duplicating the resource thereby permitting multiple modules to update the stack at the same time as long as the pushed elements satisfy $P$.
The proof structures for these two verifications are similar to that for the verification for $\Mem$:
$\textrm{is\_stk}$ and $\textrm{is\_bag}$ are defined similarly as the points-to predicate using standard authoritative PCMs;
the module-local relational invariant for the former says
each list in the pool is matched with the corresponding list stored in the module-local resource in terms of \textrm{is\_stk};
and that for the latter says
all the elements of each list in the pool satisfy the corresponding property stored in the module-local resource in terms of \textrm{is\_bag}.

Benefits of such gradual abstraction for $\Stack$ are two-fold.
First, we can achieve separation of concerns via gradual abstraction.
For example, in the first abstraction for $\Stack$,
the verification focused on abstracting linked lists into mathematical lists
without thinking about providing useful specifications to the client,
while in the second abstractions,
the verifications focused on providing such specifications based on mathematical lists.
Second, gradual abstraction increases reusability of verification results.
For example, for $\Stack$,
we essentially reused the first verification result turning linked lists into mathematical lists,
in the two verifications providing different specifications to the client.
Also note that in final abstractions obtained by applying the spec erasure theorem either with the $\textrm{is\_stk}$ specification or with the $\textrm{is\_bag}$ specification, the result of the first abstraction, $A^1_\Stack$, is reused to provide the \kwcontext{} behaviors (\ie those arising when invoked by contexts),
which is based on mathematical lists instead of linked lists.

\subsection{Abstracting function arguments and return values}
\label{sec:advanced-args}

\begin{figure}[t]
$I_\Echo$ := \hfill\mbox{}\\
\begin{minipage}[t]{0.33\textwidth}
\begin{Verbatim}[commandchars=\\\{\},codes={\catcode`$=3},fontsize=\small]
[\kwmodule Echo]
\kwdef echo([]\unwrapU{}) $\equiv$
  \kwvar stk := Stack.new();
  Echo.input(stk);
  Echo.output(stk)
\end{Verbatim}
\end{minipage}%
\begin{minipage}[t]{0.36\textwidth}
\begin{Verbatim}[commandchars=\\\{\},codes={\catcode`$=3},fontsize=\small]
\kwdef input([stk: \vval{}]\unwrapU{}) $\equiv$
  \kwvar v := IO.getint();
  \kwif (v == \exitval) \kwthen 0
  \kwelse \{ Stack.push(stk, v);
         Echo.input(stk) \}
\end{Verbatim}
\end{minipage}%
\begin{minipage}[t]{0.31\textwidth}
\begin{Verbatim}[commandchars=\\\{\},codes={\catcode`$=3},fontsize=\small]
\kwdef output([stk: \vval{}]\unwrapU{}) $\equiv$
  \kwvar v := Stack.pop(stk);
  \kwif (v == \exitval) \kwthen 0
  \kwelse \{ IO.putint(v);
         Echo.output(stk) \}
\end{Verbatim}
\end{minipage}
\\[1mm]
$A_\Echo$ := \hfill\mbox{}\\[1mm]
\begin{minipage}{0.265\textwidth}
\begin{Verbatim}[commandchars=\\\{\},codes={\catcode`$=3},fontsize=\small]
[\kwmodule Echo]
\kwdef echo = 
 \kwfriend,\kwcontext([]\unwrapU{}) $\equiv$
  \kwvar stk:\unwrapN{} list \vint
    := Echo.input([]);
  Echo.output(stk)
\end{Verbatim}
\end{minipage}%
\begin{minipage}{0.355\textwidth}
\begin{Verbatim}[commandchars=\\\{\},codes={\catcode`$=3},fontsize=\small]
\kwdef input =
 \kwfriend(stk:\unwrapN{} list \vint) $\equiv$
  \kwvar v:\unwrapU{} \vint := $\code{IO.getint()}$;
  \kwif (v == \exitval) \kwthen stk
  \kwelse Echo.input(v::stk)
 \kwcontext(\_) $\equiv$ \kwub
\end{Verbatim}
\end{minipage}%
\begin{minipage}{0.3\textwidth}
\begin{Verbatim}[commandchars=\\\{\},codes={\catcode`$=3},fontsize=\small]
\kwdef output =
 \kwfriend(stk:\unwrapN{} list \vint) $\equiv$
  \kwmatch stk \kwwith | [] => ()
  | hd::tl => IO.putint(hd);
              Echo.output(tl) \kwend
 \kwcontext(\_) $\equiv$ \kwub
\end{Verbatim}
\end{minipage}%
\hfill\mbox{}
\\[2mm]
$\mires_\Echo$ := $\munit$\hfill\mbox{}
\\
\begin{minipage}[t]{\textwidth}
\small$\begin{array}{@{}l@{}l@{\,}l@{\;}l@{}}
S_\Echo := \{
&
\code{Echo.echo}{:}
&
\snone,
&
\\&
\code{Echo.input}{:}
&
\forall \handle\!:\!\vptr.
&
\curlybracket{\lambda\,\argp\,\argv\,\argo.\; \exists \lst:\vlist\;\vint.\; \rplift{\argo = \otop \lland \argp = \upcast{[\handle]} \lland \argv = \upcast{\lst}} * \isestack{\handle}{\lst}}
\\&
&&
\curlybracket{\lambda\,\retp\,\retv.\;\exists \lst:\vlist\;\vint.\; \rplift{\retp \in \upcast{\vval} \lland \retv = \upcast{\lst}} * \isestack{\handle}{\lst}},
\\&
\code{Echo.output}{:}
&
\forall \handle\!:\!\vptr.
&
\curlybracket{\lambda\,\argp\,\argv\,\argo.\; \exists \lst:\vlist\;\vint.\; \rplift{\argo = \otop \lland \argp = \upcast{[\handle]} \lland \argv = \upcast{\lst}} * \isestack{\handle}{\lst}}
\\&
&&
\curlybracket{\lambda\,\retp\,\retv.\;\exists \lst:\vlist\;\vint.\; \rplift{\retp \in \upcast{\vval} \lland \retv = \upcast{\lst}} * \isestack{\handle}{\lst}}
\hfill\}
\end{array}$
\end{minipage}
\\
where $\isestack{\handle}{\lst} = \rplift{\lstnz(\lst)} * \isstack{\handle}{\lst}$
\hfill\mbox{}
\myhrule
\caption{An implementation and its abspec for the module \Echo{}.}
\label{fig:echo}
\end{figure}



Since friends are enforced to respect each other's specification by the spec erasure theorem,
it would make sense to abstract even function arguments and return values among the friends.
Indeed, in \CREMS, such abstraction can be easily achieved.

To see this, we consider the example given in \Cref{fig:echo}.
In the implementation $I_\Echo$\footnote{
  It shows the semantics of an \imp program, omitting the downcast and upcast around function calls.
},
\code{Echo.echo()}
%
%
%
%
%
%
%
%
%
%
%
%
creates a new stack, \code{stk};
invokes \code{Echo.input(stk)}, which repeatedly gets an integer via \code{IO.getint} and pushes it into \code{stk} until getting \code{\exitval};
and then invokes \code{Echo.output(stk)}, which repeatedly pops an integer from \code{stk} and outputs it via \code{IO.putint} until the stack is empty
(\ie \code{\exitval} is returned).
The pre-abstraction $A_\Echo$ directly uses mathematical lists instead of using the module $\Stack$.
Concretely, the \kwfriend{} definitions of \code{input} and \code{output} take a mathematical list as an argument, instead of a \vval{} value.
The argument \code{(stk:\unwrapN{} list \vint)} denotes that the \anyty{} value given as an argument is downcast to \code{list \vint}
and if it fails, triggers \kwnb{}, which \emph{guarantees} that every friend invokes them with a mathematical list.
On the other hand, we will treat the $\IO$ module as an unrestricted context (\ie specified as $\Snone$)
and thus trigger \kwub{} if the downcast from an \anyty{} value given by $\code{IO.getint()}$ to $\vint$ fails,
which is denoted by \code{\kwvar{} v:\unwrapU{} \vint{} := $\code{IO.getint()}$}.

It is important to note that the \kwcontext{} definitions of \code{input} and \code{output} immediately trigger \kwub{},
which specifies that contexts are \emph{not allowed} to invoke them
because they are intended to be such \emph{internal} functions whose arbitrary invocation may interfere the behavior of the module $\Echo$.
On the other hand, \code{echo} has the same definition for \kwfriend{} and \kwcontext{} with $\snone$,
which specifies that it can be freely invoked by both friends and contexts since it does not interfere the behavior of $\Echo$.

Now we see how we support the abstraction of argument and return values of \code{input} and \code{output} from \vval{} to \code{\vlist \vint}.
For this, \crems adds an extra argument $\argv$, called \emph{abstract argument}, and an extra return value $\retv$, called \emph{abstract return value}, to the specifications.
Then we can specify relationship between concrete ones and abstract ones in the pre and post conditions.
For example, the specifications of \code{input} and \code{output} say that the concrete argument $\argp$ contains a handle $\handle$,
the abstract one $\argv$ a list $\lst$, and they satisfy $\isestack{\handle}{\lst}$ (defined as $\rplift{\lstnz(\lst)} * \isstack{\handle}{\lst}$)
stating that $\lst$ only has non-zero elements and $\handle$ is a handle for a stack containing $\lst$;
and similarly for concrete and abstract return values.

Then we can define the abspec semantics to give an illusion of passing abstract values as follows.
When we make a call to $f$ with an abstract value (\eg \code{Echo.input(v::stk)}),
we \emph{choose} a concrete value, \emph{guarantee} that they are related as specified in the precondition of $f$,
and pass the concrete value as a real argument.
Conversely, in the \kwfriend{} definition of $f$,
we first receive a concrete value as a real argument,
and then \emph{take} an abstract value, \emph{assume} that they are related as specified in the precondition of $f$
and pass the abstract value as an argument to the \kwfriend{} definition.
The abspec semantics also does similarly for return values.
From these constructions, it follows that the spec erasure theorem can soundly eliminate concrete values and directly pass abstract ones in the final abstractions.

For verification of $\Echo$, we use $S^\twoA_\Stack$ as a specification for the module $\Stack$ and prove
\begin{equation}
  I_\Echo \lectx \spcmod{S}{(A_\Echo,\mires_\Echo)}{S_\Echo} \label{cr:echo}
\end{equation}
for any $S \sqsupseteq S_\Echo \cup S^\twoA_\Stack \cup \Snone$ with $\Snone$ for the module $\IO$.
The proof is quite straightforward:
the local simulation argument directly follows from the precondition of each function without using any module-local invariant or resource.
Note that this proof essentially involves cyclic reasoning since \code{Echo.input} and \code{Echo.output} make recursive calls;
however, it does not require any special treatment.

Now we compose all the verification results so far as follows.
By applying the spec erasure theorem to \Cref{cr:mem,cr:stack1}, and to \Cref{cr:stack2A,cr:echo}, we have
\[
\begin{array}{lcl}
  I_\Mem \link I_\Stack
  &\lectx&
  \toabs{A_\Mem}{\setofz{\Mem,\Stack}} \link \toabs{A^1_\Stack}{\setofz{\Mem,\Stack}}
  \\
  \toabs{A^1_\Stack}{\setofz{\Mem,\Stack}} \link I_\Echo
  &\lectx&
  \toabs{A^2_\Stack}{\setofz{\Stack,\Echo}} \link \toabs{A_\Echo}{\setofz{\Stack,\Echo}}
\end{array}
\]
Then by horizontal and vertical compositionality of CR, we have
\[
I_\Mem \link I_\Stack \link I_\Echo
\lectx
\toabs{A_\Mem}{\setofz{\Mem,\Stack}} \link \toabs{A^2_\Stack}{\setofz{\Stack,\Echo}} \link \toabs{A_\Echo}{\setofz{\Stack,\Echo}}
\]
Furthermore, the following three CRs hold trivially by exploiting \kwnb{} in \kwfriend{} and \kwub{} in \kwcontext{}.
\[
\toabs{A_\Mem}{\setofz{\Mem,\Stack}} \!\lectx\! I_\Mem  \ ,\ \ \ 
\toabs{A^2_\Stack}{\setofz{\Stack,\Echo}} \!\lectx\! \toabs{A^1_\Stack}{\setofz{}}  \ ,\ \ \ 
\toabs{A_\Echo}{\setofz{\Stack,\Echo}} \!\lectx\! \toabs{A_\Echo}{\setofz{\Echo}}
\]
Therefore, we can further simplify the top-level abstraction as follows.
\[
I_\Mem \link I_\Stack \link I_\Echo
\lectx
I_\Mem \link \toabs{A^1_\Stack}{\setofz{}} \link \toabs{A_\Echo}{\setofz{\Echo}}
\]

We conclude the sequence of examples shown so far with a few remarks.
First, the module $\IO$ is a part of the context, so that it can be arbitrary.
For example, it is completely valid for $\code{\IO.getint}$ and $\code{\IO.putint}$ to make mutually recursive calls to $\code{\Echo.echo}$
although calls to $\code{\Echo.input}$ or $\code{\Echo.output}$ by $\IO$ will trigger \kwub{}.
Second, it is possible to leave useful assumptions and guarantees in the final abstractions, which may be necessary or helpful for further abstractions.
For example, \code{\kwguarantee(pool handle == \none)} in $A^1_\Stack$ can be seen as such a guarantee.
As a further example, in $A_\Echo$, we can also insert \code{\kwassume(is\_prime(v))} after the call $\code{IO.getint()}$
and \code{\kwguarantee(is\_prime(hd))} before the call $\code{IO.putint(hd)}$.
Third, the definition $\snone$ (given in $S^1_\Stack$ of \Cref{fig:stack1}) says that it must be an impure call and
the concrete and abstract arguments (also return values) coincide, which is needed to prove that every \ems function semantics satisfies $\snone$.
Finally, \code{Echo.echo()} may terminate or not depending on the behavior of the $\IO$ module, which may even behave nondeterministically,
and thus its termination may be nondeterministic as well.
However, this does not cause any problem in \crems because we are not proving termination or non-termination for \code{Echo};
rather we guarantee preservation of termination: if the abstraction terminates, so does the implementation.

\subsection{Reasoning about function pointers}
\label{sec:advanced-higher}

\begin{figure}[t]
$I_\RP$ :=
\begin{minipage}[t]{0.4\textwidth}
\begin{Verbatim}[commandchars=\\\{\},codes={\catcode`$=3},fontsize=\small]
[\kwmodule RP]
\kwdef repeat([f:\vptr{}, n:\vint{}, m:\vint{}]\unwrapU{}) $\equiv$
  \kwif n $\leq$ 0 \kwthen m
  \kwelse \{ \kwvar v := (*f)(m);
         RP.repeat(f, n-1, v) \}
\end{Verbatim}
\end{minipage}%
\hfill%
\begin{minipage}[t]{0.402\textwidth}
$I_\SC$ :=
\begin{minipage}[t]{0.25\textwidth}  
\begin{Verbatim}[commandchars=\\\{\},codes={\catcode`$=3},fontsize=\small]
[\kwmodule SC]
\kwdef succ([m:\vint{}]\unwrapU{}) $\equiv$ m + 1
\end{Verbatim}
\end{minipage}\\
$I_\AD$ :=
\begin{minipage}[t]{0.25\textwidth}  
\begin{Verbatim}[commandchars=\\\{\},codes={\catcode`$=3},fontsize=\small]
[\kwmodule AD]
\kwdef add([n:\vint{}, m:\vint{}]\unwrapU{}) $\equiv$
  RP.repeat(&SC.succ, n, m)
\end{Verbatim}
\end{minipage}%
\end{minipage}
\\[1mm]
\begin{minipage}[t]{0.4\textwidth}
$A_\RP$ :=
\begin{minipage}[t]{0.25\textwidth}  
\begin{Verbatim}[commandchars=\\\{\},codes={\catcode`$=3},fontsize=\small]
[\kwmodule RP]
\kwdef repeat = \kwfriend(\_) $\equiv$ \kwnb \kwcontext \kwub
\end{Verbatim}
\end{minipage}\\
$A_\SC$ :=
\begin{minipage}[t]{0.25\textwidth}  
\begin{Verbatim}[commandchars=\\\{\},codes={\catcode`$=3},fontsize=\small]
[\kwmodule SC]
\kwdef succ = \kwfriend(\_) $\equiv$ \kwnb \kwcontext \kwub
\end{Verbatim}
\end{minipage}
\end{minipage}%
\hfill%
\begin{minipage}[t]{0.41\textwidth}
$A_\AD$ :=
\begin{minipage}[t]{0.25\textwidth}  
\begin{Verbatim}[commandchars=\\\{\},codes={\catcode`$=3},fontsize=\small]
[\kwmodule AD]
\kwdef add = \kwfriend,\kwcontext
 ([n:\vint{}, m:\vint{}]\unwrapU{}) $\equiv$
  \kwassume(n >= 0); n + m
\end{Verbatim}
\end{minipage}%
\end{minipage}
\\[1mm]
\small$\begin{array}{@{}l@{}}
H_\RP(\Sf) \!:=\! \{ \code{RP.repeat}:
\forall (f, n, m, \fsem):\vptr \times \vint \times \vint \times (\vint \rightarrow \vint).\;
\\
\hspace*{3.8pc} \{ \lambda\,\argp\,\_\,\argo.\; \rplift{\argp = \upcast{[f, n, m]} \lland n \ge 0 \lland \argo \ge \opure\;(\omega+n) \lland{}
\\
\hspace*{6.5pc}
\Sf \sqsupseteq \{ \code{*}f: \forall m\!:\!\vint, \curlybracket{\lambda\,\argp\,\_\,\argo.\, \rplift{\argp = \upcast{[m]} \land \argo = \opure\;\omega}} \ \curlybracket{\lambda\,\retp\,\_.\, \rplift{\retp = \upcast(\fsem (m))}} \} } \}
\\
\hspace*{3.8pc} \curlybracket{\lambda\,\retp\,\_.\; \rplift{\retp = \upcast{({\fsem}^n(m))}}} \hfill \}
\\  
S_\SC := \{ \code{SC.succ}: \forall m:\vint.\; \curlybracket{\lambda\,\argp\,\_\,\argo.\; \rplift{\argp = \upcast{[m]} \lland \argo = \opure\;\_}} \ \curlybracket{\lambda\,\retp\,\_.\; \rplift{\retp = \upcast(m + 1)}}  \}
\\
S_\AD := \{ \code{AD.add}: \snone \}
\end{array}$
\myhrule
\caption{Implementations and their specifications for the modules \RP, \SC, \AD}
\label{fig:repeat}
\end{figure}






In this section, we present a general pattern for doing higher-order reasoning in \crems without requiring any special support.
For this, consider the simple example given in \Cref{fig:repeat}.
The function $\code{\RP.repeat(f,n,m)}$ in $I_\RP$ recursively apply \code{*f}, \code{n} times, to \code{m},
where \code{*f} is the function pointed to by the pointer value \code{f}.
The definitions in $I_\SC$ and $I_\AD$ are straightforward to understand except that
\code{\&\SC.succ} is the pointer value pointing to the function \code{\SC.succ}.
For $\RP$ and $\SC$, we give the pure pre-abstractions $A_\RP$ and $A_\SC$,
where we maximally simplified them for presentation purposes.
For $\AD$, the pre-abstractions $A_\AD$ turns the call to \code{\RP.repeat} into the native addition with the non-negativity assumption about the first argument.

To specify \code{\RP.repeat}, we essentially need to embed expected specifications for argument functions \code{f} inside the specification of \code{\RP.repeat}.
Directly supporting this would make the definition of specification more involved since we need to solve a recursive equation to define it.
Although such an equation could be solved by employing the step-indexing technique,
here we propose a more elementary solution that does not introduce any cyclic definition.

Now we see how to do it.
First, we give a higher-order specification $H_\RP$ to the module $\RP$, given in \Cref{fig:repeat},
which is given as a function from specifications to specifications.
We can understand that the input specification $\Sf$ includes the specifications for all the functions that are passed to \code{\RP.repeat} by friends.
Then $H_\RP(\Sf)$ gives a specification for $\RP$ that only allow those functions in $\Sf$ to be given as an argument to \code{\RP.repeat}.
With this intuition, we see the definition of $H_\RP(\Sf)$:
for arguments $f,n,m$ and a mathematical function $\fsem$,
we require $\Sf$ to include the expected specification for \code{*$f$} saying that \code{*$f$} is pure with measure $\opure\;\omega$
and returns $\fsem(m)$ for any argument $m$.
Here $\omega$ is the smallest ordinal bigger than every natural number and thus we can allow \code{*$f$} to have any \emph{finite} recursion depth.
Also we require \code{\RP.repeat} to be pure with measure at least $\opure\;(\omega+n)$\footnote{
  $d \ge \opure\;o$ is defined as $\exists o' \ge o.\; d = \opure\;o'$.
}
because \code{\RP.repeat} makes recursive calls with depth $n$ followed by \mbox{a call to \code{*$f$}}.

Then we can easily verify $\RP$: for any $\Sf$ (\ie no restriction for $f$) and any $S \sqsupseteq (\Sf \cup H_\RP(\Sf))$
(since \code{\RP.repeat} makes a call to \code{*$f$} and itself), we prove
\[
I_\RP \lectx \spcmod{S}{(A_\RP,\munit)}{H_\RP(\Sf)}
\]
Also, we can easily verify $\SC$: for any $S$, we prove
\[
I_\SC \lectx \spcmod{S}{(A_\SC,\munit)}{S_\SC}
\]
Then, we can easily verify $\AD$: for any $\Sf \sqsupseteq S_\SC$ (since \code{\SC.succ} is passed to \code{\RP.repeat})
and any $S \sqsupseteq H_\RP(\Sf)$ (since \code{\AD.add} makes a call to \code{\RP.repeat}), we prove
\[
I_\AD \lectx \spcmod{S}{(A_\AD,\munit)}{S_\AD}
\]

Finally, we can instantiate the above CRs with $\Sf = S_\SC$ and $S = H_\RP(S_\SC) \cup S_\SC \cup S_\AD$
and apply the spec erasure theorem to them as follows:
\[
\begin{array}{@{}l@{~}l@{\ \ }l@{}}
  &
  I_\RP \link I_\SC \link I_\AD
  \\
  \lectx&
  \spcmod{S}{(A_\RP,\munit)}{H_\RP(S_\SC)} \!\link\! \spcmod{S}{(A_\SC,\munit)}{S_\SC} \!\link\! \spcmod{S}{(A_\AD,\munit)}{S_\AD}
  & \text{(by compositionality of CR)}
  \\
  \lectx&
  \toabs{A_\RP}{\setofz{\RP,\SC,\AD}} \link \toabs{A_\SC}{\setofz{\RP,\SC,\AD}} \link \toabs{A_\AD}{\setofz{\RP,\SC,\AD}}
  & \text{(by spec erasure thm.)}
\end{array}
\]

As a more advanced example, we also verified Landin's knot~\cite{irislecture}, which can be found in our Coq development~\cite{supplementary}.
Since we do not know of any practical higher-order example for which our approach fails,
we believe it is general enough in practice.


\section{Formal Definitions of \CREMS}
\label{sec:formal}

\begin{figure}[t]
\small
\raggedright

$\myset{X}{cond} \defeq
\text{if } cond \text{ holds, then } X \text{ else } \emptyset
\qquad\qquad
\trm{fundef}(E) \defeq \tyany \rightarrow \trm{itree}\;E\;\tyany$
\\[1mm]
\begin{tabbing}
$
\EventE(X)$ \= $\defeq$
\= $\{\chooseE \} \uplus \{ \takeE \} \uplus
\myset{\curlybracket{\syscallE{} \; \fn \; args \;|\; \fn \in \code{string},\, args \in \tyany}}
      {X = \tyany}
$\\
$
\EmsE(X)$ \> $\defeq$
 \> $\EventE(X) \uplus \myset{\curlybracket{\callE \; \fn \; args \;|\; \fn \in \code{string},\, args \in \tyany}}{X=\tyany} \uplus {}$\\
 \> \> $\myset{\curlybracket{\getE}}{X=\tyany} \uplus
 \myset{\curlybracket{\putE\;a \;|\; a \in \tyany}}{X=\unitset} \uplus
 \myset{\curlybracket{\getNameE{}}}{X=\code{string}} $\\
$
\ems$ \> $\defeq$ \> $\curlybracket{ (\code{name}, \code{init}, \code{funs}) \in \code{string} \times \tyany \times
  (\code{string} \fpfn{} \trm{fundef}(\EmsE))}
$
\end{tabbing}
\begin{tabbing}
$\SpcE(X)$ \= $\defeq \EmsE(X) \uplus \myset{\curlybracket{\ApcE}}{X=\unitset} $\\
$\gen$ \> $\defeq \curlybracket{ (\code{name}, \code{init}, \code{funs}) \in \code{string} \times \tyany \times
  (\code{string} \fpfn{} (\trm{fundef}(\SpcE) \times \trm{fundef}(\SpcE)))} $
\end{tabbing}
\vspace*{1mm}
\begin{tabbing}
PCM \defeq
$\{$\=$(\monoid,\; \madd,\; \wdef,\; \munit) \in (Set
\times (\monoid \rightarrow \monoid \rightarrow \monoid) \times \mathcal{P}(\monoid) \times
\monoid) \; | \; $\madd$\, \text{is commutative and associative,} $ \\
 \> $\munit$ is an identity element, $\wdef(\munit)$ holds, and $\wdef$ is monotone with respect to addition. \}
\end{tabbing}
\begin{tabbing}
$\rProp_\Sigma \;$ \= $\defeq$ \= $\Sigma \rightarrow \textbf{Prop}$ \ \ \ \ for $\Sigma \in \textrm{PCM}$\\
$\trm{Spec}_\Sigma \;$ \= $\defeq$ \= $\setofz{(\Meta, \cnd) \;|\; \Meta \in \trm{Type} \land
     \cnd \in \Meta \rightarrow (\tyany \rightarrow \tyany \rightarrow
     \option\;\ordinal \rightarrow \rProp_\Sigma) \times (\tyany \rightarrow \tyany \rightarrow \rProp_\Sigma)} $\\
$\trm{Specs}_\Sigma$ \> $\defeq$ \> $\code{string} \fpfn{} \trm{Spec}_\Sigma $
\end{tabbing}
\begin{tabbing}
$s_{1}$ \= $\sqsupseteq s_{0} \in \trm{Spec}\ \ $  \= $\defeq$ \=
$\forall \meta_{0} \in (s_{0}.\Meta)\,.\;
\exists \meta_{1} \in (s_{1}.\Meta)\,.\; \exists (P_{0}, Q_{0}) = s_{0}.\cnd(\meta_{0})\,.\; \exists (P_{1}, Q_{1}) = s_{1}.\cnd(\meta_{1})\,.\;$\\
\> \> \>
$(\forall\,\argp\,\argv\,\argo.\; (P_{0}\,\argp\,\argv\,\argo) \vdash \upd{} (P_{1}\,\argp\,\argv\,\argo))
\land
(\forall\,\retp\,\retv.\; (Q_{1}\,\retp\,\retv) \vdash \upd{} (Q_{0}\,\retp\,\retv)) $\\
$S_{1}$ \> $\sqsupseteq S_{0} \in \trm{Specs}\ \ $ \> $\defeq \forall \fn\in\code{string}\,.\; \forall s_{0} \in \trm{Spec}\,.\;
S_{0}(\fn) = \some\;s_{0} \implies
\exists s_{1} \in \trm{Spec},\; S_{1}(\fn) = \some\;s_{1} \land s_{1} \sqsupseteq s_{0}
$
\end{tabbing}
\vspace*{1mm}
$\emsmod \defeq \LD \times (\LD \rightarrow \ems) \qquad
\emsmods \defeq \code{list } \emsmod \qquad
\llink \in \emsmods \rightarrow \emsmods \rightarrow \emsmods \defeq \code{append} $\\
$\trm{Trace} \defcoind
\curlybracket{e :: tr \;|\; e \in \trm{ObsEvent}, tr \in \trm{Trace} } \uplus
\curlybracket{\trm{Term } v \;|\; v \in \tyany{}} \uplus \curlybracket{\trm{Diverge}} \uplus
\curlybracket{\trm{Error}} \uplus \curlybracket{\trm{Partial}}
 $\\
$\text{Beh} \in \emsmods \rightarrow \mathbb{P}\trm{(Trace)} \defeq ... $\\
$ M_\iside \lectx M_\aside \in \emsmods \defeq \forall\, C \in \emsmods\,.\; \beh{C \llink M_\iside} \subseteq \beh{C \llink M_\aside} $
\myhrule
\caption{Formal definitions of \CREMS}
\label{formal:coredefs}
\end{figure}


In this section, we present formal definitions, key properties, and the embedding of abspecs and abstraction into \ems.
%
First of all, our Coq formalization largely relies on interaction trees~\cite{liyao:itree}.
Intuitively, the interaction tree $\code{itree}\;E\;T$ for an event type $E$ (defining a set of events $E(X)$ for each set $X$)
and a return type $T$ can be understood as a small-step operational semantics that can take a silent step, terminate with a value in $T$,
or trigger an event in $E(X)$ for any set $X$, in which case its continuation nondeterministically receives each value in $X$.
We enjoy two benefits of interaction trees: $(i)$ they provide useful combinators, which made our various constructions straightforward,
and $(ii)$ they can be extracted to executable programs in OCaml.
In particular, all the language constructs shown in the examples so far are simply combinators for itrees.
The key combinator we use is the \emph{interpretation} function with states in $ST$, which has type:
\[
\code{itree}\;E\;T \rightarrow (\forall X.\; E(X) \rightarrow ST \rightarrow \code{itree}\;E'\;(X\times ST)) \rightarrow ST \rightarrow \code{itree}\;E'\;(T\times ST)
\]
where we use the notation $t[e_1 \mapsto t_1,\ldots,e_n \mapsto t_n]$ to denote the interpretation of
the events $e_i$ to the (state-indexed) itree $t_i$ in the itree $t$
and implicitly drop identical interpretations essentially mapping $e$ to itself
and also drop the state component when $ST = ()$.

\Cref{formal:coredefs} shows the formal definitions of \crems.
First, we define two notations $\myset{X}{cond}$ for conditional construction of a set
and $\trm{fundef}(E)$ for semantics of a function that takes a value in \tyany{} and
executes by possibly triggering events in the event type $E$.
We first define the primitive events $\EventE$ consisting of $\chooseE$ and $\takeE$ for any type $X$,
and $\syscallE$ for triggering observable events such as system calls
that pass and receive an \anyty{} value;
and the events $\EmsE$ extends $\EventE$ with $\callE$ for making a function call,
$\getE$ and $\putE$ for reading from and writing to the module local state,
and $\getNameE$ for getting the caller's module name.
Then \ems{} for semantics of a module
is defined as the set of triples consisting of
a module's name, an initial module-local state and function semantics triggering events in $\EmsE$ for (a finite set of) module functions.
The events $\SpcE$ extends $\EmsE$ with $\ApcE$ for triggering an IPC,
and \gen{} for pre-abstraction of a module is defined similarly as \ems{}
except that \gen{} has a pair of function semantics (for \kwfriend{} and \kwcontext{})
triggering events in $\SpcE$ for each module function.

PCM, the set of PCMs, is defined in a standard way~\cite{irisgroundup},
where the predicate $\wdef$ indicates whether a resource is defined or not.
A specification (for a function) in $\trm{Spec}_\Sigma$, parameterized by a global PCM $\Sigma$,
consists of a set $\Meta$ over which the universal quantifier in the specification quantifies,
and a condition $\cnd$ that, given a value $\meta \in \Meta$, gives a pair of pre and post conditions.
A precondition takes concrete and abstract arguments with a measure and gives a resource proposition, which is a predicate on $\Sigma$;
and similarly for a postcondition but without measure.
A collection of specifications in $\trm{Specs}_\Sigma$ is a finite map from function names to function specifications.
We also define the strengthening relation $\sqsupseteq$ between specifications,
which generalizes the simple inclusion relation following \iris~\cite{irisgroundup}.


$\emsmod$ gives a notion of code for a single module (\ie before loading),
which is parameterized by a notion of loading data $\LD$,
which happens to be required to form a PCM to combine loading data from all modules and express consistency between them.
A \emph{module code} consists of its own loading data in $\LD$ and a function in $\LD \rightarrow \ems$
that, given the global loading data gathered from all the modules, returns its module semantics.
A \emph{modules code} in $\emsmods$ is simply a list of module codes, which forms basic units in contextual refinement,
and linking $\link$ between them is the list append.


Then we coinductively define the set of traces, Trace.
A trace is a finite or infinite sequence of observable events in \trm{ObsEvents},
defined as $\setofz{(\syscallE{} \; \fn \; args, r) \;|\; \fn \in \code{string},\, args, r \in \tyany}$,
possibly ended with one of the four cases:
$(i)$ normal termination with an \anyty{} value,
$(ii)$ divergence without producing any observable events,
$(iii)$ fatal error,
or $(iv)$ partial termination.
The notion of partial termination can be intuitively understood as terminating the execution at the user's will such as pressing \texttt{Ctrl+C},
which is dual to fatal error (\ie termination due to the program's fault).
This will be clarified in the definition of $\textrm{Beh}$
that gives the set of those traces that can possibly arise when loading and executing a given modules code.
The definition (omitted for brevity; see our Coq development~\cite{supplementary} for details)
is defined as usual except that
$(i)$ the partial termination, \textrm{Partial}, may occur nondeterministically at any point during execution
(capturing that the user can stop the program at any time),
and
$(ii)$ triggering a \textrm{Choose} (or \textrm{Take}) event is interpreted as taking the union (or intersection) of the behaviors of all possible continuations.
Note that triggering \kwub{} (\ie \emph{taking} from the empty set) can produce all possible traces,
which is dual to the interpretation of triggering \kwnb{} (\ie \emph{choosing} from the empty set)
that can only immediately terminate with \textrm{Partial} without any other possible traces
(capturing that there is \emph{no behavior} caused by the program after \kwnb{} is triggered).
Finally, we define \emph{contextual refinement} $\lectx$ between modules codes as usual.

In \crems, we establish CR using a standard simulation technique that allows
module-local relational invariants, \simrel{}, to depend on Kripke-style possible worlds equipped with a preorder ($\simle$).
Then, the (coinductively-defined) greatest simulation relation $\lesssim_w$ at a given world $w$
relates target states (\ie the implementation side) and source states (\ie the abspec side),
both of which consist of a module-local state and an itree\footnote{
  We also support the stuttering index using our ordinal library, but omit it here for brevity.
}.
Two notable cases are the return and call cases shown below:
in the former one has to prove that the invariant \simrel{} holds at the current or a future world;
and relying on that, in the latter, one can assume the invariant holds after every function call.
\[
\small
\begin{array}{c@{\quad}c@{\quad}c@{\quad}c}
\infer{ w \simle w' \qquad \simrel \, w' \, st_\iside \, st_\aside}
      {(st_\iside, \kw{ret}\;r) \lesssim_{w} (st_\aside, \kw{ret}\;r)} \qquad
      \infer{w \simle w' \qquad  \simrel \, w' \, st_\iside \, st_\aside \\\\ \forall r, w'', st'_\iside, st'_\aside \,.\ 
        w' \simle w'' \land \simrel \, w'' \, st'_\iside \, st'_\aside \Rightarrow
      (st'_\iside, K_\iside[r]) \lesssim_{w} (st'_\aside, K_\aside[r])
}
      {(st_\iside, \kwvar{}\; r \code{:=} \kw{call} \; f \;x; \; K_\iside[r]) \lesssim_{w}
       (st_\aside, \kwvar{}\; r \code{:=} \kw{call} \; f \;x; \; K_\aside[r])
}
\end{array}
\]

\begin{figure}[t]
\hfill
\begin{minipage}{0.525\textwidth}

  {
\small
$\toabs{A}{N} \defeq$ $( A.\code{name},\; A.\code{init},\;$ $\lambda \fn.\; \code{\transA} (N, A.\code{funs} \; \fn) )$
}%

\begin{Verbatim}[commandchars=\\\{\},codes={\catcode`$=3\catcode`_=8},fontsize=\small]
\trm{\transA}$(N, (frd, ctx))$: \trm{fundef}(\EmsE) $\defeq{}\lambda{}\argp.$
  \kwvar mn := \kwgetcaller{}();
  \kwif(mn $\in$ $N$) $frd$($\argp$)[IPC $\mapsto$ \kwskip]
  \kwelse   \hspace{0.20cm}    $ctx$($\argp$)[IPC $\mapsto$ \kwskip]
\end{Verbatim}

{
\small
$\spcmod{S}{A, \mires}{S_{a}} \defeq$ $( A.\code{name},\; (A.\code{init}, \upcast{\mires}),\;$\\
\-\hspace{2cm}  $\lambda\, \fn.\; \code{\transAS}(S,\;A.\code{funs} \; \fn, \; S_{a}\;\fn) )$
}%

\begin{Verbatim}[commandchars=\\\{\},codes={\catcode`$=3\catcode`_=8},fontsize=\small]
\trm{\transAS}$(S, (frd, ctx), s)$: \trm{fundef}(\EmsE) $\defeq{}\lambda{}\argp.$
  \kwvar is\_friend := \kwtake(bool);
  \kwif(is\_friend) $\textrm{\AS{}Fun}(S, s, frd)(x)$
  \kwelse          $\textrm{\AS{}Fun}(S, \snone,\,ctx)(x)$
\end{Verbatim}

\begin{Verbatim}[commandchars=\\\{\},codes={\catcode`$=3\catcode`_=8},fontsize=\small]
\trm{\AS{}Fun}($S$, $s$, fun: \trm{fundef}(\SpcE)) $\defeq$ $\lambda{}\argp.$
  \kwvar ($\Meta$, PQ) := $s$;
  \kwvar \meta{} := \kwtake(\Meta); \kwvar (P, Q) := PQ(\meta{});
  \kwvar ($\argv$, $\argo$, \frm{}) := \ASSUME(P, $\argp$, $\munit$);
  \kwmatch $\argo$ \kwwith
  | None => \kwvar ($\retv$, \frm{}) :=
    \AS{}Body($S$, $\argo$, \frm{}, fun($\argv$))
  | Some o =>
    \kwvar (\_, \frm{}) := \AS{}IPC($S$, $\argo$, \frm{});
    \kwvar $\retv$ := \kwchoose(\tyany)  \kwend;
  \kwvar ($\retp$, \_, \_) := \GUARANTEE(Q, $\retv$, \frm{});
  $\retp$
\end{Verbatim}

\begin{Verbatim}[commandchars=\\\{\},codes={\catcode`$=3\catcode`_=8},fontsize=\small]
\AS{}Body($S$, $\argo$, \frm{}, body) $\defeq$ body[
 \callE fn $\argv$ $\mapsto$ $\lambda{}\frm{}.$ \trm{\AS{}Call}($S$,$\argo$,\frm{},fn,$\argv$),
   \hspace{0.59cm} \ApcE $\mapsto$ $\lambda{}\frm{}.$ \trm{\AS{}IPC}($S$,$\argo$,\frm{})]
\end{Verbatim}

\end{minipage}%
\hfill
\begin{minipage}{0.470\textwidth}

\begin{Verbatim}[commandchars=\\\{\},codes={\catcode`$=3\catcode`_=8},fontsize=\small]
\trm{\AS{}Call}($S$, $\argo$, \frm{}, fn, $\argv$) $\defeq$
  \kwvar ($\Meta$, PQ) := $S$(fn);
  \kwvar \meta := \kwchoose($\Meta$); \kwvar (P, Q) := PQ(\meta);
  \kwvar ($\argp$,$\argo$',\kwfp) := \GUARANTEE(P,$\argv$,\frm);
  \kwguarantee($\argo$' $\sqsubset \argo$);
  \kwvar $\retp$ := \kwcall fn $\argp$;
  \kwvar ($\retv$, \_, \frm) := \ASSUME(Q, r, \kwfp);
  ($\retv$, \frm{})
\trm{\AS{}IPC}($S$, $\argo$, \frm{}) $\defeq$
  \kwvar i := \kwchoose(\ordinal);
  \kwwhile(\kwchoose(bool))
    \kwvar (fn, $\argv$) := \kwchoose(\fname $\times$ \anyv);
    (\_, \frm{}) := \trm{\AS{}Call}($S$,$\argo$,\frm{},fn,$\argv$);
    i := \kwchoose(\{i' $\in$ \ordinal | i' < i\})
  ((), \frm{})

\color{red}\ASSUME(Cond, $\xorv$, \kwfp) $\equiv$ \{
\color{red}  \kwvar ($\xorv_{a}$,$\argo$,\res,\frm) := \kwtake(\_);
\color{red}  \kwvar (\kwmp, \_) := \kw{get};
\color{red}  \kwassume(Cond $\xorv$ $\xorv_{a}$ $\argo$ \res);
\color{red}  \kwassume$(\mval(\res+\kwfp+\kwmp+\frm{}))$;
\color{red}  ($\xorv_{a}$, $\argo$, \frm) \}
\color{blue}\GUARANTEE(Cond, $\xorv_{a}$, \frm) $\equiv$ \{
\color{blue}  \kwvar ($\xorv$,$\argo$',\res,\kwfp,\kwmp) := \kwchoose(\_);
\color{blue}  \kwguarantee(Cond $\xorv$ $\xorv_{a}$ $\argo$' \res);
\color{blue}  \kwguarantee$(\mval(\res+\kwfp+\kwmp+\frm{}))$;
\color{blue}  \kwvar (\_, st) := \kw{get}; \kw{put}(\kwmp, st);
\color{blue}  ($\xorv$, $\argo$', \kwfp) \}
\end{Verbatim}
\end{minipage}
\myhrule
\caption{Translations of abspecs and pre-abstractions into \ems.}
\label{formal:embedding2}
\end{figure}

\Cref{formal:embedding2} formally presents our translation of abspecs into \ems
and that of pre-abstractions into \ems (\ie abstractions).
These translations are done in the same way as we have explained except that, as we mentioned before,
instead of requiring frame-preserving updates we allow to update local resources as long as they are consistent with the frame resource given at the latest interaction point.
Also note that we used the same macros $\ASSUME$ and $\GUARANTEE$ for both pre and post conditions although the latter lacks the measure parameter.
Here we implicitly cast postconditions to the type of preconditions by making them to ignore the measure parameter.

Now we present our core theorems. The most important one -- spec erasure theorem -- is
already presented in \Cref{thm:spec-erasure}.
Then we present adequacy of the simulation relation.
\begin{theorem} [Adequacy]\label{thm:adequacy} For a given pair of module $M_\iside$ and $M_\aside$,
  a possible world $\world$ equipped with $\simle$, and a module-local relational invariant $\simrel$,
  if each pair of functions with the same name for $M_\iside$ and $M_\aside$
  are related by the simulation relation $\lesssim$
  for any argument and any module-local states satisfying $I$,
  then we have the contextual refinement between them:
\[
  [M_\iside] \lectx [M_\aside]
\]
\end{theorem}

We also use the following strengthening theorem.
\begin{theorem} [Strengthening]\label{thm:weakening}
  For any $S, S', A, \mires, S_A$, the following holds:
\[
S' \sqsupseteq S \implies \spcmod{S}{A, \mires}{S_A} \lectx \spcmod{S'}{A,\mires}{S_A}
\]
\end{theorem}

Finally, we briefly discuss how to define the abstraction \Safe mentioned in \Cref{sec:introduction}.
The module $\Safe(\names, \names')$ defines each function in $\names'$ whose
semantics is simply defined to nondeterministically invoke an arbitrary function in $\names$ with arbitrary arguments any number of times (even infinitely many).
Then we have the following theorem.
\begin{theorem} [Safety]\label{thm:erasure-safe}
  For $\names = \names_{1} \uplus \ldots \uplus \names_{n}$, $\Safe(\names, \names_{1}) \llink \ldots \llink \Safe(\names, \names_{n})$ is safe
  (\ie never produces an \trm{Error}).
\end{theorem}

\section{Proofmode}
\label{sec:proofmode}

\crems proof mode consists of two modes supporting simulation reasoning and separation logic reasoning.

\subsection{Simulation Reasoning}

\myparagraph{Reduction Tactics}
We establish CR using a simulation technique,
which relates an implementation state and an abspec state\footnote{
  In fact, our simulation tactics generally work for relating \emph{any} two \ems{} semantics.
}.
To take steps,
we provide reduction tactics that reduces
the current itree (either in implementation or in abspec) into the head normal form
by, \eg converting $(i \mbind j) \mbind k$ into $i \mbind (j \mcat k)$.
Also, we made the tactics extensible:
when embedding a new language with new effects and a new handler H into \ems{},
we can register new reduction lemmas about H
so that they are used by the reduction tactics
(\eg converting $H(i \mbind j)$ into $H i \mbind H j$).

\myparagraph{Simulation Tactics}
We provide tactics that automate common parts of simulation reasoning.
Specifically, the tactics repeatedly apply the rules given below (\ie converting the conclusion into the premise),
where we omitted the state component and the world index in the simulation relation for brevity.
Note that the \dashbox{dash-boxed} rules are \emph{complete} (\ie the premise and conclusion are equivalent) while the others not.
Then we provide three tactics:
\kw{steps} automatically applies complete rules as many as possible,
\kw{force\_i} applies any available rule once in the implementation side,
and 
\kw{force\_a} does the same for the abspec side.
\[
\scriptsize
\begin{array}{c@{\quad}c@{\quad}c@{\quad}c}
\infer{\exists x \in X.~T_\iside \lesssim K_\aside[x]}
      {T_\iside \lesssim \code{x := \kwchoose($X$); $K_\aside[\code{x}]$}} &
\dashbox{
\infer{\forall x\in X.~ T_\iside \lesssim K_\aside[x]}
      {T_\iside \lesssim \code{x := \kwtake($X$); $K_\aside[\code{x}]$}}} &
\infer{P \lland (T_\iside \lesssim T_\aside)}
      {T_\iside \lesssim \code{\kwguarantee($P$); $T_\aside$}} &
\dashbox{
\infer{P \implies T_\iside \lesssim T_\aside}
      {T_\iside \lesssim \code{\kwassume($P$); $T_\aside$}}}
\\

\span\\

\dashbox{
\infer{\forall x\in X.~ K_\iside[x] \lesssim T_\aside}
      {\code{x := \kwchoose($X$); $K_\iside[\code{x}]$} \lesssim T_\aside}} &
\infer{\exists x\in X.~ K_\iside[x] \lesssim T_\aside}
      {\code{x := \kwtake($X$); $K_\iside[\code{x}]$} \lesssim T_\aside} &
\dashbox{
\infer{P \implies T_\iside \lesssim T_\aside}
      {\code{\kwguarantee(P); $T_\iside$} \lesssim T_\aside}} &
\infer{P \lland (T_\iside \lesssim T_\aside)}
      {\code{\kwassume(P); $T_\iside$} \lesssim T_\aside}
\\
\end{array}
\]

\subsection{Separation Logic Reasoning}

\[
\forall x.\; P \;|\; \exists x.\; P \;|\; \lc P \rc \;|\;
P \land Q \;|\; P \lor Q \;|\; P \Rightarrow Q \;|\;
P * Q \;|\; P \wand Q \;|\;
\ownGhost{}{a: \Sigma} \;|\; \upd{} P \;|\; P \vdash Q
\]
We support the SL tactics provided by IPM (Iris Proof Mode).
For this, we define the standard \iris connectives shown above on $\rProp$, instead of $\mathbf{iProp}$ of \iris with step indices,
and prove the lemmas that IPM (Iris Proof Mode) requires, which allows us to use IPM when proving logical entailment.
Note that we omitted the later modality ($\latert$) because we do not use step-indexing.

\myparagraph{At Interaction Points}
We have three tactics -- \kw{slinit}, \kw{slcall}, and \kw{slret} -- for each interaction
points. \kw{slinit} is used at the beginning of a function and it moves the precondition (in $\rProp$)
into Coq hypothesis (\code{current\_rProps}).
\kw{slcall} is used when calling a function ---
where source have its call decorated with \code{\AS\_call} (\Cref{formal:embedding2}) and
target not. It asks the user to specify a subset of \code{current\_rProps} that will be
consumed when proving (with IPM) both (i) callee's precondition and (ii) the relational
invariant. After that, the user proceeds the proof with fresh \trm{rProp}s satisfying (i)
callee's postcondition and (ii) the relational invariant. In other words, while the mechanism
is implemented with various \kwassume{}/\kwguarantee{}s in low-level (\eg \code{\AS\_call}), we
give a high-level interface with these tactics. \kw{slret} is used at the end of the function
and is similar to \kw{slcall}.


\myparagraph{In Between} We also support operations on \code{current\_rProps} in the middle
of the simulation proof. Notably, we support (i) eliminating connectives like $\forall,
\exists, \lor, \land, *, \upd{}, \lc \rc$, (ii) \kw{Assert} tactic that consumes specified
\trm{rProp}s to establish specified goal (using IPM), and (iii) \kw{OwnV} tactic that gives
$\wdef(\mires)$ for given $\ownGhost{}{\mires}$.  It is worth noting that, despite such
functionalities, it suffices to put updates and validity checking only at the interaction
points (as in \Cref{sec:overview:separation}), not in between.




\section{Imp and Compilation to CompCert}\label{sec:imp}

The \imp{} language, extended from \code{Imp} presented in \cite{liyao:itree}, has the following syntax and its semantics is defined in \ems.
\[
\begin{array}{@{}l@{\;}l@{\;}l@{\;}l@{}}
  \multicolumn{4}{@{}l@{}}{x, f, g \in String \qquad f\!p \in Pointer} \\
  e &\in Expr & ::= & x \;|\; v:\mathbb{Z} \;|\; e_{1} \: == \: e_{2} \;|\;
  e_{1} < e_{2} \;|\; e_{1} + e_{2} \;|\; e_{1} - e_{2} \;|\; e_{1} \times e_{2} \\

  s &\in Stmt & ::= & \code{skip} \;|\; x := e \;|\; s_{1};s_{2} \;|\; \code{if} \: (e) \: \code{then} \: \{s_{1}\} \: \code{else} \: \{s_{2}\} \;|\;
  x = f(e_{1},...,e_{n}) \;|\; x = f\!p(e_{1},...,e_{n}) \;|\; \\
  & & & x = \code{\&}g \;|\; x = \code{malloc}(e) \;|\; \code{free}(e) \;|\;
  x = \code{load}(e) \;|\; \code{store}(e_{1},e_{2}) \;|\; x = \code{cmp}(e_{1},e_{2})
\end{array}
\]
We have developed a verified compiler\footnote{
  As a simple solution to resolve a subtle mismatch between CompCert's memory model and our simplified one,
  we compile the \code{free} instruction to \code{skip}.}
from \imp to Csharpminor of CompCert~\cite{CompCert},
which is then composed with CompCert to give a compiler $\mathcal{C}$ from \imp{} to CompCert's assembly.
\begin{theorem} [Separate Compilation Correctness]\label{imp:correctness}
  Let $(I_1,Asm_1), \ldots, (I_n,Asm_n)$ be pairs of \imp and Asm modules
  such that $\mathcal{C}(I_i) = \some(Asm_i)$ for all $i$.
  Then, the following holds:
  \[
  \beh{Asm_1 \plink \cdots \plink Asm_n}\footnote{
    We cast CompCert's events into \code{Obs} events in \ems.
  }
  \subseteq \beh{\memimpl \llink I_1 \llink \cdots
    \llink I_n} ~.
  \]
\end{theorem}
Note that CompCert assemblies are linked with the \emph{syntactic linking} operator (\plink)
and we followed the lightweight verification approach of \cite{kang:scc}.

\section{Evaluation}
\label{sec:discussion}

\begin{wraptable}{r}{5.0cm}
\vspace{-5mm}
\centering
\begin{tabu}{l @{\;}|@{\;} l @{\;}|@{\;} l}
Portion & Def & Proof \\
\hline
\ems & 2584 & 3025 \\
SL-specific & 5709 & 3632 \\
\imp-specific & 3770 & 4129 \\ 
Examples & 3082 & 3054 \\ 
Coq libraries & 2689 & 850 \\ 
\hline
Total & 17834 & 14690 \\
\end{tabu}
\label{table:evaluation}
\vspace{-3mm}
\end{wraptable}

The right table shows the SLOC of the whole development (counted by \code{coqwc}).
\ems{} and its meta-theory amount to 5609 SLOC, SL-specific theory (including \gen{}, proof
mode, translations, and spec erasure theorem) to 9341 SLOC, \imp{}-specific theory (including
syntax, semantics, its compiler, and compiler verification) to 7899 SLOC, verification
examples (Cannon, Mem, Stack, Echo, Repeat, and Landin's knot) to 6136 SLOC, and general
purpose Coq libraries to 3539 SLOC. In total, our Coq definitions amount to 17834 SLOC and
proofs to 14690 SLOC.

The proof effort for verifying each example in \crems is split into two:
$(i)$ that for reasoning about SL entailment, which would be comparable to that in \iris because we use IPM and prove essentially similar goals,
and
$(ii)$ that for simulation reasoning, which was quite straightforward in all the examples thanks to our tactics.

Vertical compositionality of CR (\ie gradual abstraction) also simplified the proof of meta theory of \crems.
For instance, the spec erasure theorem (\Cref{thm:spec-erasure}) is established by
transitively composing six contextual refinements, two major ones of which are
$(i)$ removing \ASSUME{} and \GUARANTEE{} while concretizing the measure information
and $(ii)$ removing pure calls by proving their termination using the measure information.


All the examples in the paper are extracted to OCaml programs, and we indeed found bugs by testing before verification.
For extraction, all the events are handled inside Coq except for \EventE{},
which are handled by special handlers written in OCaml.
Specifically, we wrote a few handlers doing IO for $\syscallE$
and a handler for $\chooseE$ and $\takeE$, which asks the user for a nondeterministic choice
(currently only supports \vint{}).
For testing, we extracted both implementations and abstractions, executed them and compared the results.
Interestingly, we found two mis-downcast bugs in the abstraction of \Echo{} by testing before verification.
Also we can extract and run even the abspecs although it would introduce so much nondeterminism.





\section{Related Work} 
\label{sec:related}

There have been many works in proving safety and refinement of programs in various directions but occasionally with certain restrictions.
Abstraction logic can be seen as a unifying theory, based on elementary mechanisms, that can subsume most of the works without such restrictions.
We will discuss those works and their restrictions if any.



\myparagraph{Specifications as programs}

Refinement calculus~\cite{back2012refinement} understands Hoare-style specifications as programs and
provides refinement between them, which enjoys fully compositionality as in CR.
\cite{jeremie:lics, jeremie:thesis} recently made advances in this line of research,
where they also employ dual-nondeterminism and algebraic effects (similar to interaction trees but without extraction) like \crems.
However, unlike \crems, they do not support SL-style specifications.
\hide{
\myparagraph{Dual nondeterminism} The concept is already present in the literature\cite{
tyrrell2006lattice}, but few of
them used it for writing specifications, and none connected it with SL. The one that is
closest to our approach is that of \cite{jeremie:lics, jeremie:thesis}, where dual
non-determinism is formalized using domain theory, used in writing specifications, and such
specifications are verified with \yj{TODO}.  Also, they employ algebraic effects which
corresponds to interaction trees, but they do not support extraction to OCaml. \yj{COMPLETELY REWRITE THIS PARAGRAPH}
}




\myparagraph{Separation Logics}

First of all, compared to the state-of-the-art separation logics such as Iris and VST, \CREMS does not support concurrency yet
although we plan to extend \crems to support it following their approaches.


There have been works based on separation logics that go beyond safety such as CaReSL\cite{caresl} and ReLoC\cite{reloc}.
Like \crems, they establish contextual refinement using SL but in a restricted setting that
does not allow transformations essentially relying on logical specifications of external modules.

Using \iris, \cite{sandbox} establishes guarantee of desired
properties on observable traces (\ie a sequence of system calls),
instead of safety guarantee, in the presence of unverified contexts,
but in a restricted setting that does not allow the contexts to invoke system calls.


\hide{
As discussed (\Cref{sec:introduction}), standard version of SL
has several limitations and there are various works each extending with different directions.


Notable works that go beyond mere safety are CaReSL\cite{caresl} and ReLoC\cite{reloc}
but they have different focus with us. On one hand, they have advantages of
supporting concurrency and general higher-order features, utilizing step-index. On the other
hand, while they do establish CR, they also inherit the limitations of it
(\Cref{sec:introduction}) that they cannot use SL specifications for shared
states. Consequently, their verification examples are restricted to abstracting non-shared
states (references that have not been leaked to outside). Nonetheless, such abstraction is
not sound if an arbitrary context illegally accesses such location, so they rule out such
context by typing.
}





\myparagraph{Contextual Refinement}

Certified Abstraction Layers (CAL)\cite{gu:dscal,ccal} proved effectiveness of contextual refinement in large scale verification
by verifying a realistic operating system.
Compared to \crems, although it supports concurrency, CAL is limited in a few aspects.
For example, CAL does not support SL-style specifications and thus does not allow implementations to use shared resources across modules.
Also it does not allow mutual recursion between modules.


\section{Conclusion and Future Work}


We present a comprehensive theory combining the benefits of contextual refinement and separation logic, together with practical tools,
using the key idea of \kwchoose{} and \kwtake{} that gives an illusion of passing any information to anyone without involving physical operations.
As future works, we plan to extend \CREMS to support concurrency,
and also develop testing tools that can efficiently find bugs that breaks desired contextual refinement,
which may also give a certain level of confidence without verification.


\hide{
\myparagraph{Future Work} We are interested in (i) extending the framework to concurrent
setting, and (ii) supporting automated testing seriously. For the former, we believe that our
most fundamental ideas will scale to concurrent setting. Furthermore, it would be interesting to
adopt the ideas from \cite{liang:pp} to specify and verify progress properties with CR. For
the latter, we are interested in adopting SMT-solver\cite{???} or property-based
testing\cite{???}. We believe that gradual abstraction will play an important role in doing
so, as noticed in \cite{ARMADA}.
}





\begin{acks}                            
  This material is based upon work supported by the
  \grantsponsor{GS100000001}{National Science
    Foundation}{http://dx.doi.org/10.13039/100000001} under Grant
  No.~\grantnum{GS100000001}{nnnnnnn} and Grant
  No.~\grantnum{GS100000001}{mmmmmmm}.  Any opinions, findings, and
  conclusions or recommendations expressed in this material are those
  of the author and do not necessarily reflect the views of the
  National Science Foundation.
\end{acks}

\newpage
\balance
\bibliography{references}


\begin{thebibliography}{26}


\ifx \showCODEN    \undefined \def \showCODEN     #1{\unskip}     \fi
\ifx \showDOI      \undefined \def \showDOI       #1{#1}\fi
\ifx \showISBNx    \undefined \def \showISBNx     #1{\unskip}     \fi
\ifx \showISBNxiii \undefined \def \showISBNxiii  #1{\unskip}     \fi
\ifx \showISSN     \undefined \def \showISSN      #1{\unskip}     \fi
\ifx \showLCCN     \undefined \def \showLCCN      #1{\unskip}     \fi
\ifx \shownote     \undefined \def \shownote      #1{#1}          \fi
\ifx \showarticletitle \undefined \def \showarticletitle #1{#1}   \fi
\ifx \showURL      \undefined \def \showURL       {\relax}        \fi
\providecommand\bibfield[2]{#2}
\providecommand\bibinfo[2]{#2}
\providecommand\natexlab[1]{#1}
\providecommand\showeprint[2][]{arXiv:#2}

\bibitem[\protect\citeauthoryear{Ahmed}{Ahmed}{2006}]%
        {ahmed2006step}
\bibfield{author}{\bibinfo{person}{Amal Ahmed}.}
  \bibinfo{year}{2006}\natexlab{}.
\newblock \showarticletitle{Step-indexed syntactic logical relations for
  recursive and quantified types}. In \bibinfo{booktitle}{\emph{European
  Symposium on Programming}}. Springer, \bibinfo{pages}{69--83}.
\newblock


\bibitem[\protect\citeauthoryear{Appel}{Appel}{2011}]%
        {vst}
\bibfield{author}{\bibinfo{person}{Andrew~W. Appel}.}
  \bibinfo{year}{2011}\natexlab{}.
\newblock \showarticletitle{Verified Software Toolchain}. In
  \bibinfo{booktitle}{\emph{Proceedings of the 20th European Symposium on
  Programming}} \emph{(\bibinfo{series}{ESOP 2011})}.
\newblock


\bibitem[\protect\citeauthoryear{Author(s)}{Author(s)}{2021}]%
        {supplementary}
\bibfield{author}{\bibinfo{person}{Anonymous Author(s)}.}
  \bibinfo{year}{2021}\natexlab{}.
\newblock \bibinfo{title}{Supplementary material for this paper available to
  reviewers}.
\newblock
\newblock


\bibitem[\protect\citeauthoryear{Back and Wright}{Back and Wright}{2012}]%
        {back2012refinement}
\bibfield{author}{\bibinfo{person}{Ralph-Johan Back} {and}
  \bibinfo{person}{Joakim Wright}.} \bibinfo{year}{2012}\natexlab{}.
\newblock \bibinfo{booktitle}{\emph{Refinement calculus: a systematic
  introduction}}.
\newblock \bibinfo{publisher}{Springer Science \& Business Media}.
\newblock


\bibitem[\protect\citeauthoryear{Birkedal and Bizjak}{Birkedal and
  Bizjak}{2020}]%
        {irislecture}
\bibfield{author}{\bibinfo{person}{Lars Birkedal} {and}
  \bibinfo{person}{Ale{\v{s}} Bizjak}.} \bibinfo{year}{2020}\natexlab{}.
\newblock \bibinfo{title}{Lecture notes on iris: Higher-order concurrent
  separation logic}.
\newblock
\newblock
\urldef\tempurl%
\url{https://iris-project.org/tutorial-material.html}
\showURL{%
\tempurl}


\bibitem[\protect\citeauthoryear{Bodik, Chandra, Galenson, Kimelman, Tung,
  Barman, and Rodarmor}{Bodik et~al\mbox{.}}{2010}]%
        {bodik2010programming}
\bibfield{author}{\bibinfo{person}{Rastislav Bodik}, \bibinfo{person}{Satish
  Chandra}, \bibinfo{person}{Joel Galenson}, \bibinfo{person}{Doug Kimelman},
  \bibinfo{person}{Nicholas Tung}, \bibinfo{person}{Shaon Barman}, {and}
  \bibinfo{person}{Casey Rodarmor}.} \bibinfo{year}{2010}\natexlab{}.
\newblock \showarticletitle{Programming with angelic nondeterminism}. In
  \bibinfo{booktitle}{\emph{Proceedings of the 37th annual ACM SIGPLAN-SIGACT
  symposium on Principles of programming languages}}.
  \bibinfo{pages}{339--352}.
\newblock


\bibitem[\protect\citeauthoryear{Calcagno, O'Hearn, and Yang}{Calcagno
  et~al\mbox{.}}{2007}]%
        {calcagno2007local}
\bibfield{author}{\bibinfo{person}{Cristiano Calcagno},
  \bibinfo{person}{Peter~W O'Hearn}, {and} \bibinfo{person}{Hongseok Yang}.}
  \bibinfo{year}{2007}\natexlab{}.
\newblock \showarticletitle{Local action and abstract separation logic}. In
  \bibinfo{booktitle}{\emph{22nd Annual IEEE Symposium on Logic in Computer
  Science (LICS 2007)}}. IEEE, \bibinfo{pages}{366--378}.
\newblock


\bibitem[\protect\citeauthoryear{Frumin, Krebbers, and Birkedal}{Frumin
  et~al\mbox{.}}{2018}]%
        {reloc}
\bibfield{author}{\bibinfo{person}{Dan Frumin}, \bibinfo{person}{Robbert
  Krebbers}, {and} \bibinfo{person}{Lars Birkedal}.}
  \bibinfo{year}{2018}\natexlab{}.
\newblock \showarticletitle{ReLoC: A mechanised relational logic for
  fine-grained concurrency}. In \bibinfo{booktitle}{\emph{Proceedings of the
  33rd Annual ACM/IEEE Symposium on Logic in Computer Science}}.
  \bibinfo{pages}{442--451}.
\newblock


\bibitem[\protect\citeauthoryear{Gu, Koenig, Ramananandro, Shao, Wu, Weng,
  Zhang, and Guo}{Gu et~al\mbox{.}}{2015}]%
        {gu:dscal}
\bibfield{author}{\bibinfo{person}{Ronghui Gu},
  \bibinfo{person}{J{\'{e}}r{\'{e}}mie Koenig}, \bibinfo{person}{Tahina
  Ramananandro}, \bibinfo{person}{Zhong Shao},
  \bibinfo{person}{Xiongnan~(Newman) Wu}, \bibinfo{person}{Shu{-}Chun Weng},
  \bibinfo{person}{Haozhong Zhang}, {and} \bibinfo{person}{Yu Guo}.}
  \bibinfo{year}{2015}\natexlab{}.
\newblock \showarticletitle{Deep Specifications and Certified Abstraction
  Layers}. In \bibinfo{booktitle}{\emph{Proceedings of the 42nd {ACM}
  {SIGPLAN-SIGACT} Symposium on Principles of Programming Languages}}
  \emph{(\bibinfo{series}{POPL 2015})}.
\newblock


\bibitem[\protect\citeauthoryear{Gu, Shao, Kim, Wu, Koenig, Sj{\"o}berg, Chen,
  Costanzo, and Ramananandro}{Gu et~al\mbox{.}}{2018}]%
        {ccal}
\bibfield{author}{\bibinfo{person}{Ronghui Gu}, \bibinfo{person}{Zhong Shao},
  \bibinfo{person}{Jieung Kim}, \bibinfo{person}{Xiongnan Wu},
  \bibinfo{person}{J{\'e}r{\'e}mie Koenig}, \bibinfo{person}{Vilhelm
  Sj{\"o}berg}, \bibinfo{person}{Hao Chen}, \bibinfo{person}{David Costanzo},
  {and} \bibinfo{person}{Tahina Ramananandro}.}
  \bibinfo{year}{2018}\natexlab{}.
\newblock \showarticletitle{Certified concurrent abstraction layers}.
\newblock \bibinfo{journal}{\emph{ACM SIGPLAN Notices}} \bibinfo{volume}{53},
  \bibinfo{number}{4} (\bibinfo{year}{2018}), \bibinfo{pages}{646--661}.
\newblock


\bibitem[\protect\citeauthoryear{Hoare}{Hoare}{1969}]%
        {hoare1969logic}
\bibfield{author}{\bibinfo{person}{Charles Antony~Richard Hoare}.}
  \bibinfo{year}{1969}\natexlab{}.
\newblock \showarticletitle{An axiomatic basis for computer programming}.
\newblock \bibinfo{journal}{\emph{Commun. ACM}} \bibinfo{volume}{12},
  \bibinfo{number}{10} (\bibinfo{year}{1969}), \bibinfo{pages}{576--580}.
\newblock


\bibitem[\protect\citeauthoryear{Jung, Krebbers, Jourdan, Bizjak, Birkedal, and
  Dreyer}{Jung et~al\mbox{.}}{2018}]%
        {irisgroundup}
\bibfield{author}{\bibinfo{person}{Ralf Jung}, \bibinfo{person}{Robbert
  Krebbers}, \bibinfo{person}{Jacques-Henri Jourdan},
  \bibinfo{person}{Ale{\v{s}} Bizjak}, \bibinfo{person}{Lars Birkedal}, {and}
  \bibinfo{person}{Derek Dreyer}.} \bibinfo{year}{2018}\natexlab{}.
\newblock \showarticletitle{Iris from the ground up: A modular foundation for
  higher-order concurrent separation logic}.
\newblock \bibinfo{journal}{\emph{Journal of Functional Programming}}
  \bibinfo{volume}{28} (\bibinfo{year}{2018}).
\newblock


\bibitem[\protect\citeauthoryear{Jung, Swasey, Sieczkowski, Svendsen, Turon,
  Birkedal, and Dreyer}{Jung et~al\mbox{.}}{2015}]%
        {iris2015}
\bibfield{author}{\bibinfo{person}{Ralf Jung}, \bibinfo{person}{David Swasey},
  \bibinfo{person}{Filip Sieczkowski}, \bibinfo{person}{Kasper Svendsen},
  \bibinfo{person}{Aaron Turon}, \bibinfo{person}{Lars Birkedal}, {and}
  \bibinfo{person}{Derek Dreyer}.} \bibinfo{year}{2015}\natexlab{}.
\newblock \showarticletitle{Iris: Monoids and invariants as an orthogonal basis
  for concurrent reasoning}.
\newblock \bibinfo{journal}{\emph{ACM SIGPLAN Notices}} \bibinfo{volume}{50},
  \bibinfo{number}{1} (\bibinfo{year}{2015}), \bibinfo{pages}{637--650}.
\newblock


\bibitem[\protect\citeauthoryear{Kang, Hur, Mansky, Garbuzov, Zdancewic, and
  Vafeiadis}{Kang et~al\mbox{.}}{2015}]%
        {int2ptr}
\bibfield{author}{\bibinfo{person}{Jeehoon Kang}, \bibinfo{person}{Chung-Kil
  Hur}, \bibinfo{person}{William Mansky}, \bibinfo{person}{Dmitri Garbuzov},
  \bibinfo{person}{Steve Zdancewic}, {and} \bibinfo{person}{Viktor Vafeiadis}.}
  \bibinfo{year}{2015}\natexlab{}.
\newblock \showarticletitle{A formal C memory model supporting integer-pointer
  casts}.
\newblock \bibinfo{journal}{\emph{ACM SIGPLAN Notices}} \bibinfo{volume}{50},
  \bibinfo{number}{6} (\bibinfo{year}{2015}), \bibinfo{pages}{326--335}.
\newblock


\bibitem[\protect\citeauthoryear{Kang, Kim, Hur, Dreyer, and Vafeiadis}{Kang
  et~al\mbox{.}}{2016}]%
        {kang:scc}
\bibfield{author}{\bibinfo{person}{Jeehoon Kang}, \bibinfo{person}{Yoonseung
  Kim}, \bibinfo{person}{Chung-Kil Hur}, \bibinfo{person}{Derek Dreyer}, {and}
  \bibinfo{person}{Viktor Vafeiadis}.} \bibinfo{year}{2016}\natexlab{}.
\newblock \showarticletitle{Lightweight Verification of Separate Compilation}.
  In \bibinfo{booktitle}{\emph{Proceedings of the 43rd {ACM} {SIGPLAN-SIGACT}
  Symposium on Principles of Programming Languages}}
  \emph{(\bibinfo{series}{POPL 2016})}.
\newblock


\bibitem[\protect\citeauthoryear{Koenig}{Koenig}{2020}]%
        {jeremie:thesis}
\bibfield{author}{\bibinfo{person}{J\'{e}r\'{e}mie Koenig}.}
  \bibinfo{year}{2020}\natexlab{}.
\newblock \showarticletitle{Refinement-Based Game Semantics for Certified
  Components}.
\newblock
\urldef\tempurl%
\url{https://flint.cs.yale.edu/flint/publications/koenig-phd.pdf}
\showURL{%
\tempurl}


\bibitem[\protect\citeauthoryear{Koenig and Shao}{Koenig and Shao}{2020}]%
        {jeremie:lics}
\bibfield{author}{\bibinfo{person}{J\'{e}r\'{e}mie Koenig} {and}
  \bibinfo{person}{Zhong Shao}.} \bibinfo{year}{2020}\natexlab{}.
\newblock \showarticletitle{Refinement-Based Game Semantics for Certified
  Abstraction Layers}. In \bibinfo{booktitle}{\emph{Proceedings of the 35th
  Annual ACM/IEEE Symposium on Logic in Computer Science}} (Saarbr\"{u}cken,
  Germany) \emph{(\bibinfo{series}{LICS '20})}. \bibinfo{publisher}{Association
  for Computing Machinery}, \bibinfo{address}{New York, NY, USA},
  \bibinfo{pages}{633–647}.
\newblock
\showISBNx{9781450371049}
\urldef\tempurl%
\url{https://doi.org/10.1145/3373718.3394799}
\showDOI{\tempurl}


\bibitem[\protect\citeauthoryear{Leroy}{Leroy}{2006}]%
        {CompCert}
\bibfield{author}{\bibinfo{person}{Xavier Leroy}.}
  \bibinfo{year}{2006}\natexlab{}.
\newblock \showarticletitle{Formal Certification of a Compiler Back-end or:
  Programming a Compiler with a Proof Assistant}. In
  \bibinfo{booktitle}{\emph{Proceedings of the 33rd {ACM} {SIGPLAN-SIGACT}
  Symposium on Principles of Programming Languages}}
  \emph{(\bibinfo{series}{POPL 2006})}.
\newblock


\bibitem[\protect\citeauthoryear{O’hearn}{O’hearn}{2007}]%
        {ohearn2007csl}
\bibfield{author}{\bibinfo{person}{Peter~W O’hearn}.}
  \bibinfo{year}{2007}\natexlab{}.
\newblock \showarticletitle{Resources, concurrency, and local reasoning}.
\newblock \bibinfo{journal}{\emph{Theoretical computer science}}
  \bibinfo{volume}{375}, \bibinfo{number}{1-3} (\bibinfo{year}{2007}),
  \bibinfo{pages}{271--307}.
\newblock


\bibitem[\protect\citeauthoryear{Reynolds}{Reynolds}{2002}]%
        {reynolds2002separation}
\bibfield{author}{\bibinfo{person}{John~C Reynolds}.}
  \bibinfo{year}{2002}\natexlab{}.
\newblock \showarticletitle{Separation logic: A logic for shared mutable data
  structures}. In \bibinfo{booktitle}{\emph{Proceedings 17th Annual IEEE
  Symposium on Logic in Computer Science}}. IEEE, \bibinfo{pages}{55--74}.
\newblock


\bibitem[\protect\citeauthoryear{Sammler, Garg, Dreyer, and Litak}{Sammler
  et~al\mbox{.}}{2019}]%
        {sandbox}
\bibfield{author}{\bibinfo{person}{Michael Sammler}, \bibinfo{person}{Deepak
  Garg}, \bibinfo{person}{Derek Dreyer}, {and} \bibinfo{person}{Tadeusz
  Litak}.} \bibinfo{year}{2019}\natexlab{}.
\newblock \showarticletitle{The high-level benefits of low-level sandboxing}.
\newblock \bibinfo{journal}{\emph{Proceedings of the ACM on Programming
  Languages}} \bibinfo{volume}{4}, \bibinfo{number}{POPL}
  (\bibinfo{year}{2019}), \bibinfo{pages}{1--32}.
\newblock


\bibitem[\protect\citeauthoryear{{\v{S}}ev{\v{c}}{\'\i}k, Vafeiadis,
  Zappa~Nardelli, Jagannathan, and Sewell}{{\v{S}}ev{\v{c}}{\'\i}k
  et~al\mbox{.}}{2013}]%
        {compcerttso}
\bibfield{author}{\bibinfo{person}{Jaroslav {\v{S}}ev{\v{c}}{\'\i}k},
  \bibinfo{person}{Viktor Vafeiadis}, \bibinfo{person}{Francesco
  Zappa~Nardelli}, \bibinfo{person}{Suresh Jagannathan}, {and}
  \bibinfo{person}{Peter Sewell}.} \bibinfo{year}{2013}\natexlab{}.
\newblock \showarticletitle{CompCertTSO: A verified compiler for relaxed-memory
  concurrency}.
\newblock \bibinfo{journal}{\emph{Journal of the ACM (JACM)}}
  \bibinfo{volume}{60}, \bibinfo{number}{3} (\bibinfo{year}{2013}),
  \bibinfo{pages}{1--50}.
\newblock


\bibitem[\protect\citeauthoryear{{The Coq Development Team}}{{The Coq
  Development Team}}{2021}]%
        {coq}
\bibfield{author}{\bibinfo{person}{{The Coq Development Team}}.}
  \bibinfo{year}{2021}\natexlab{}.
\newblock \bibinfo{title}{The {Coq} Proof Assistant 8.13.2 Reference Manual}.
\newblock
\newblock
\newblock
\shownote{\url{https://coq.github.io/doc/V8.13.2/refman/}}.


\bibitem[\protect\citeauthoryear{Turon, Dreyer, and Birkedal}{Turon
  et~al\mbox{.}}{2013}]%
        {caresl}
\bibfield{author}{\bibinfo{person}{Aaron Turon}, \bibinfo{person}{Derek
  Dreyer}, {and} \bibinfo{person}{Lars Birkedal}.}
  \bibinfo{year}{2013}\natexlab{}.
\newblock \showarticletitle{Unifying refinement and Hoare-style reasoning in a
  logic for higher-order concurrency}. In \bibinfo{booktitle}{\emph{Proceedings
  of the 18th ACM SIGPLAN international conference on Functional programming}}.
  \bibinfo{pages}{377--390}.
\newblock


\bibitem[\protect\citeauthoryear{Tyrrell, Morris, Butterfield, and
  Hughes}{Tyrrell et~al\mbox{.}}{2006}]%
        {tyrrell2006lattice}
\bibfield{author}{\bibinfo{person}{Malcolm Tyrrell}, \bibinfo{person}{Joseph~M
  Morris}, \bibinfo{person}{Andrew Butterfield}, {and} \bibinfo{person}{Arthur
  Hughes}.} \bibinfo{year}{2006}\natexlab{}.
\newblock \showarticletitle{A lattice-theoretic model for an algebra of
  communicating sequential processes}. In
  \bibinfo{booktitle}{\emph{International Colloquium on Theoretical Aspects of
  Computing}}. Springer, \bibinfo{pages}{123--137}.
\newblock


\bibitem[\protect\citeauthoryear{Xia, Zakowski, He, Hur, Malecha, Pierce, and
  Zdancewic}{Xia et~al\mbox{.}}{2019}]%
        {liyao:itree}
\bibfield{author}{\bibinfo{person}{Li-yao Xia}, \bibinfo{person}{Yannick
  Zakowski}, \bibinfo{person}{Paul He}, \bibinfo{person}{Chung-Kil Hur},
  \bibinfo{person}{Gregory Malecha}, \bibinfo{person}{Benjamin~C. Pierce},
  {and} \bibinfo{person}{Steve Zdancewic}.} \bibinfo{year}{2019}\natexlab{}.
\newblock \showarticletitle{Interaction Trees: Representing Recursive and
  Impure Programs in Coq}.
\newblock \bibinfo{journal}{\emph{Proc. ACM Program. Lang.}}
  \bibinfo{volume}{4}, \bibinfo{number}{POPL}, Article \bibinfo{articleno}{51}
  (\bibinfo{date}{Dec.} \bibinfo{year}{2019}), \bibinfo{numpages}{32}~pages.
\newblock
\urldef\tempurl%
\url{https://doi.org/10.1145/3371119}
\showDOI{\tempurl}


\end{thebibliography}


\end{document}